\documentclass[aps,pre,reprint,amsmath,amssymb,groupedaddress,showpacs]{revtex4-1}
\usepackage{graphicx}
\usepackage{subfigure}
\usepackage{epstopdf}
\usepackage{multirow}

\begin{document}

%\preprint{}

%\title{Scaling phenomena in failure mechanisms of load sharing complex systems}
\title{Failure mechanisms of load sharing complex systems}


%\author{}
%\thanks{}
%\affiliation{}
\author{Shahnewaz Siddique}
\email[email: ]{sids@gatech.edu}
\author{Vitali Volovoi}%
\affiliation{%
School of Aerospace Engineering, Georgia Institute of Technology, Atlanta, GA 30332 USA
}%


\date{\today}

\begin{abstract}
We investigate the failure mechanisms of load sharing complex systems. The system is composed of multiple nodes or components whose failures are determined based on the interaction of their respective strengths and loads (or capacity and demand respectively) as well as the ability of a component to share its load with its neighbors when needed. We focus on two distinct mechanisms to model the interaction between components' strengths and loads. The failure mechanisms of these two models demonstrate temporal scaling phenomena, phase transitions and multiple distinct failure modes excited by extremal dynamics. For critical ranges of parameters the models demonstrate power law and exponential failure patterns. We identify the similarities and differences between the two mechanisms and the implications of our results to the failure mechanisms of complex systems in the real world.

\end{abstract}

\pacs{89.75.Da, 89.75.Fb, 89.75.Kd, 05.10.-a, 02.70.-c }
%\keywords{}

\maketitle

%\section{Introduction and Model}

In the last decade a significant body of research has accumulated in the study of complex systems, their structure and dynamics \cite{newman:1,*boccaletti:1}. Static robustness in terms of node removals have been explored in \cite{albert:1,*crucitti:2,*crucitti:3}. But most real networks undergo dynamic failures where the failure of a single or multiple nodes might trigger cascades of failure through the network. Dynamical redistribution of flow have been considered in different real world networks: power grids \cite{kinney:1, *crucitti:1}, air transportation networks \cite{lacasa:1}, and communication networks \cite{martino:1}.

Many physical systems fail as their capacity or strength degrades over time under constant load or load increases over time as strength remains fixed. For example, loss of strength phenomena is observed in stress-rupture or creep rupture \cite{mahesh:1}, tire wear, and level of fluid in a hydraulic system \cite{lemoine:1}; whereas load buildup is commonly considered in fiber-bundle models on complex systems \cite{moreno:1,kim:1}. Failure occurs when component load is greater than its strength. Component failure due to overloading is a serious threat in networks: a single component failure and its subsequent load redistribution can trigger cascades of failures through the network, ultimately bringing down the entire system \cite{moreno:1, kinney:1}.

On the other hand, many communication and transportation systems exhibit congestion phenomena as data or customer traffic density increases beyond certain thresholds. Congestion or jamming phenomena for critical values of traffic density has been demonstrated in models of transportation \cite{nagel:1,*nagel:2,*nagel:3} and communication networks \cite{ohira:1,sole:1,valverde:1}. Traffic flow models for air transportation systems have been explored in \cite{lacasa:1, roy:1, sridhar:1}. Congestion in one part of the network has the effect of rerouting traffic to other parts of the network resulting in slowing down or clogging traffic in the entire system.

In this work we explore two different models of interaction between component strength and load to understand the failure mechanisms of complex systems. The first one is a Loss of Strength (LOS) model where components lose strength over time following prescribed rules. The second one is a Customer Service (CS) model where component demand is modeled through customer or data traffic arrival rate. For both models we investigate the strength-load interaction both at and below critical loading levels \cite{moreno:1,moreno:2}. At critical loading levels and above the entire system abruptly collapses, which we refer to as the critical state.

First, we describe the general system setup. We implement both models on two different network topologies: individual components organized in a \(n\) \(\times\) \(n\) lattice or a scale-free (SF) network of $n^2$ components following a power law degree distribution $P(k) \sim Ak^{-\gamma}$, with exponents $2 < \gamma < 3$. The SF network is constructed using the Barab\'asi-Albert (BA) algorithm \cite{barabasi:1,*barabasi:2}. The BA SF model is a growth and preferential attachment algorithm where at each iteration step a new node is attached to ``$m$'' existing nodes in the network, where $m$ is a constant input parameter. At the end of the iteration steps, a scale-free network of average degree  $\langle k\rangle = 2m$ is obtained. We generate BA SF networks for $m \in (2,4,6)$ which results in average degree $\langle k\rangle \in (4,8,12)$. These choices of $\langle k\rangle$ cover a range of communication, biological and social networks in the real world \cite{boccaletti:1}. In both network topologies each component can be in one of three possible modes: fully operational, stressed, and failed. We denote by \((i,j)\) the location of a component in the lattice. For the scale-free configuration we number the components from 1 to $n^2$.  Next we describe the LOS model in the lattice configuration.

%\section{The LOS model}
%\section{Failure models}

For the LOS model on a lattice topology each component is initialized with a specific strength \(S_{ij}\). Component loads \(L_{ij}\) are initialized with the same value and during the simulation are not exogenously varied. If \(L_{ij} \leq \eta S_{ij}\), where $\eta \in (0,1)$ is a parameter to control strength degradation threshold, then the component is fully operational and strength does not degrade. If \(\eta S_{ij} < L_{ij} \leq S_{ij}\) then the component at \((i,j)\) is considered stressed and loss of strength takes place over time. We consider deterministic loss of strength for components \cite{lemoine:1}. The components strength degradation follows the relationship $S^{t}_{ij}= -\alpha t  + S^{t'}_{ij}$ where $\alpha$ is the strength degradation rate parameter, $t$ is the time and $S^{t'}_{ij}$ denotes the strength at time $t=t'$ when component LOS commences. If \(L_{ij} > S_{ij}\) then the component fails and the load is redistributed equally to the components immediate neighbors in the system. Once a component fails it is removed from the network.

The objective of the \(S_{ij}\) and \(L_{ij}\) initializations is to capture the interaction dynamics of component strength and load. The simulations work in the following way, first all component loads are set to a specific value $L_{ij}=L$ where $L \in [0.5 .. 4]$. During a simulation $L$ is not exogenously varied. For each load setting $L$, 30,000 Monte-Carlo simulations are carried out and \(S_{ij}\) is reinitialized for each simulation. To generate a mix of strong and weak components, \(S_{ij}\) is initialized from a real uniform distribution $\mathcal{U}[6,10]$. To initiate LOS dynamics, for each simulation 4-5 components are initialized to a stressed mode $L_{ij} > \eta S_{ij}$ where deterministic loss of strength takes place. As \(L\) is steadily increased, we arrive at critical ranges of \(L\) where interaction between components \(L_{ij}\) and \(S_{ij}\) trigger LOS dynamics and load redistributions for increasingly greater number of components. This allows us to capture the failure mechanisms of the system. In the simulations $t=k\Delta t$ with $\Delta t = 0.1$, $k=1..500$, $n=12$, $\eta=0.7$ and $\alpha = 0.2$. The components at the boundary are initialized to very high strength to prevent failure. Since boundary components do not fail we do not need to deal with their load redistributions.

In the SF network case each component has neighbors following a power-law degree distribution. The simulation initializations for LOS SF network model is the same as the lattice configuration except with $n^2=100$ and for each simulation we generate a BA SF network. Also by construction all components for the LOS model on a SF network have neighbors so special handling of boundary components is not necessary.

%\section{The CS model}

Next we describe the CS model on a lattice configuration. For the CS model we have taken a ``Eulerian'' \cite{lamb:1} point of view for component flow dynamics as opposed to the standard ``Lagrangian'' point of view of our references. Component demand is modeled as a customer or data arrival rate $\lambda$. Although traffic in real communication networks is non-Poissonian \cite{sole:1, paxson:1}, as a first step we follow \cite{roy:1,ohira:1} and model customer demand as a Poisson process with rate $\lambda$. The rate $\lambda$ is the same for all components and does not vary during a simulation. Thus the system is in effect subjected to a uniformly distributed globally varying load. Component capacity is modeled through a fixed customer departure rate $\gamma_{ij}$.  In addition, each component possesses an associated queue $q_{ij}$ for extra storage capacity. At a given time step if $\lambda \leq \gamma_{ij}$ then the component is fully operational. If $\lambda > \gamma_{ij}$ then the excess demand ($\lambda - \gamma_{ij}$) is redistributed to the fully operational neighbors of the component. Excess demand is transferred sequentially to the neighbors with the largest spare capacity $(\gamma_{ij} - \lambda) > 0$ where $(i,j)$ denotes the location of the neighbors. If component demand redistribution is successful then the component remains fully operational. The component excess demand redistribution might be partially or completely unsuccessful. In that event, the remaining excess demand is placed in the queue $q_{ij}$ for processing in the next time step. If the remaining excess demand is placed in $q_{ij}$ successfully then the component at $(i,j)$ is considered stressed. If the remaining excess demand overwhelms $q_{ij}$ then the component is considered congested (failed) as it is not able to service the traffic demand. Once a component is congested it is taken off the grid.

Similar to the LOS model, we capture the interaction between component capacity and demand for the CS model through critical ranges of $\lambda$ that trigger demand redistribution and congestion. In our simulations, component capacities $\gamma_{ij}$ are initialized by sampling from a integer uniform distribution $\mathcal{U}[8,12]$ to generate a mix of strong and weak components. We run 30,000 Monte-Carlo simulations for each integer value of $\lambda \in [5,11]$. For each simulation the system is initialized with new capacities $\gamma_{ij}$. Each simulation is run for $t=k\Delta t = 500$ time steps where $\Delta t = 1$. The queue size is set to $q=6$ and $n=12$. For our ranges of $\lambda$ the queue essentially provides components additional time to prevent failure. Boundary components have queue's set to large values to prevent component failure and avoid load redistribution. For the CS model on a SF network we have $n^2=100$ and due to circular boundary conditions special handling of boundary components is not necessary.

%\section{TF and TT distributions}
\begin{figure*}[!]
  \begin{center}
    \subfigure[$L=0.5$]{\label{fig:lostfloadpt5}\includegraphics[scale=0.25]{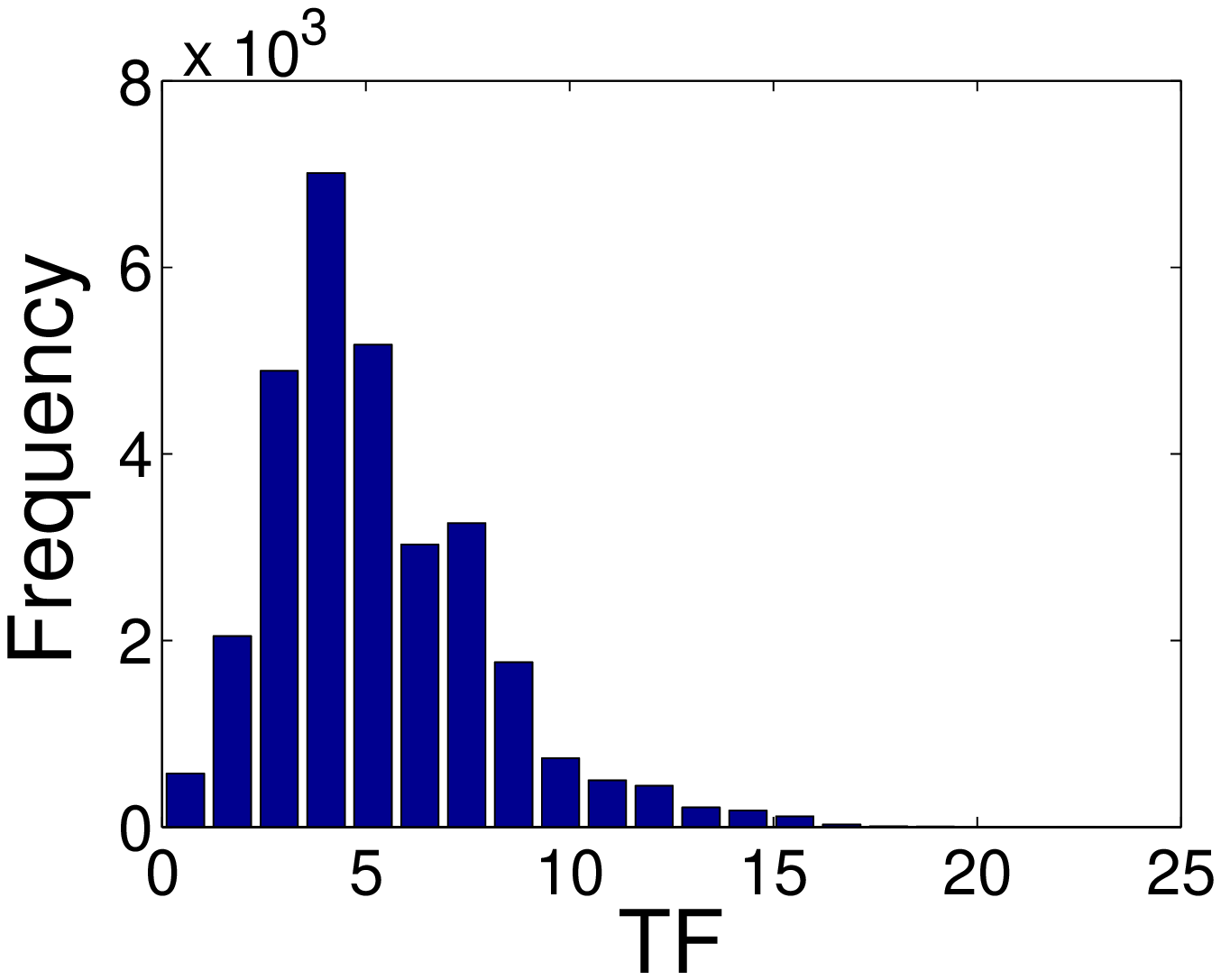}}
    \subfigure[$L=2$]{\label{fig:lostfload2}\includegraphics[scale=0.25]{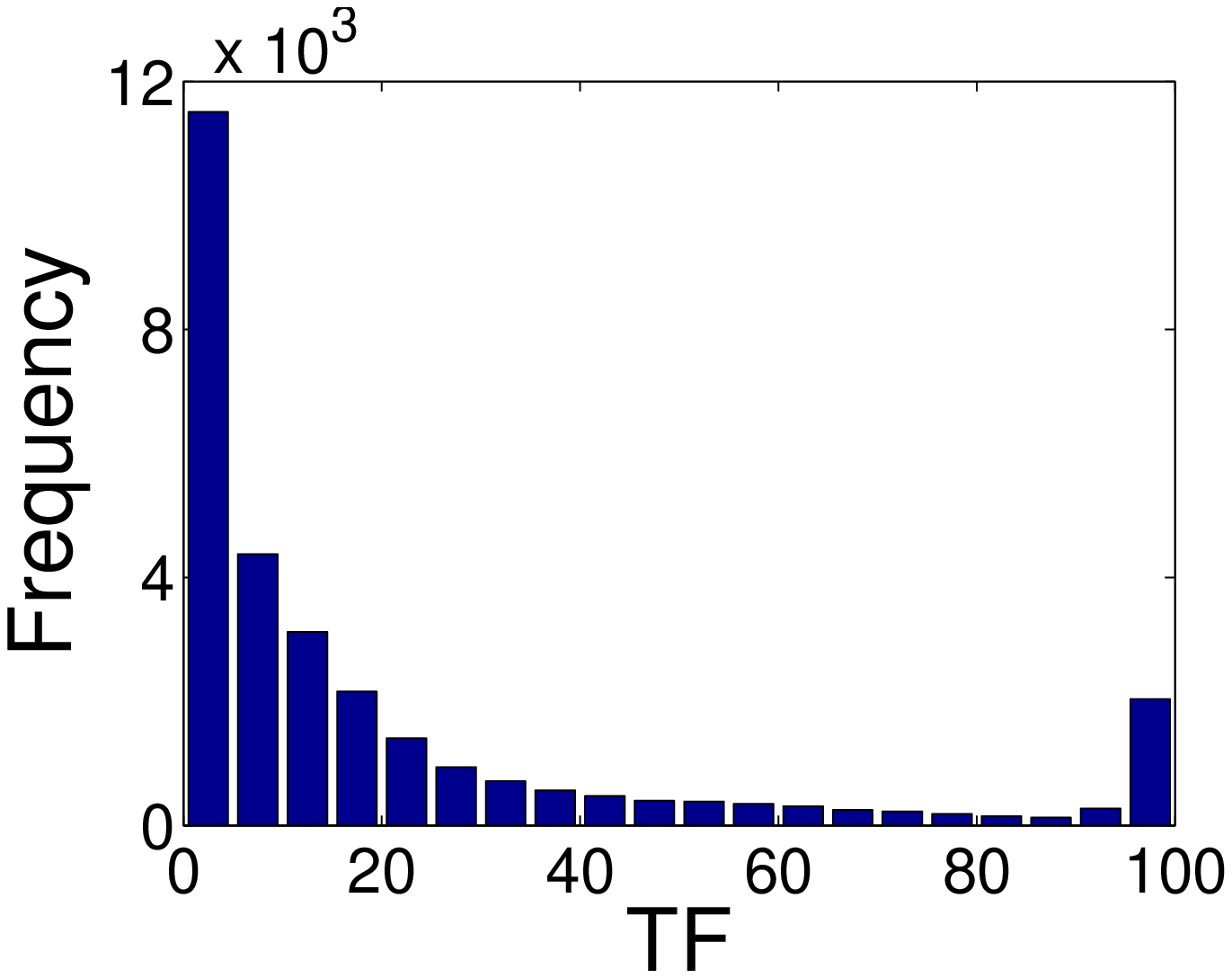}}
    \subfigure[$L=2.5$]{\label{fig:lostfload2pt5}\includegraphics[scale=0.25]{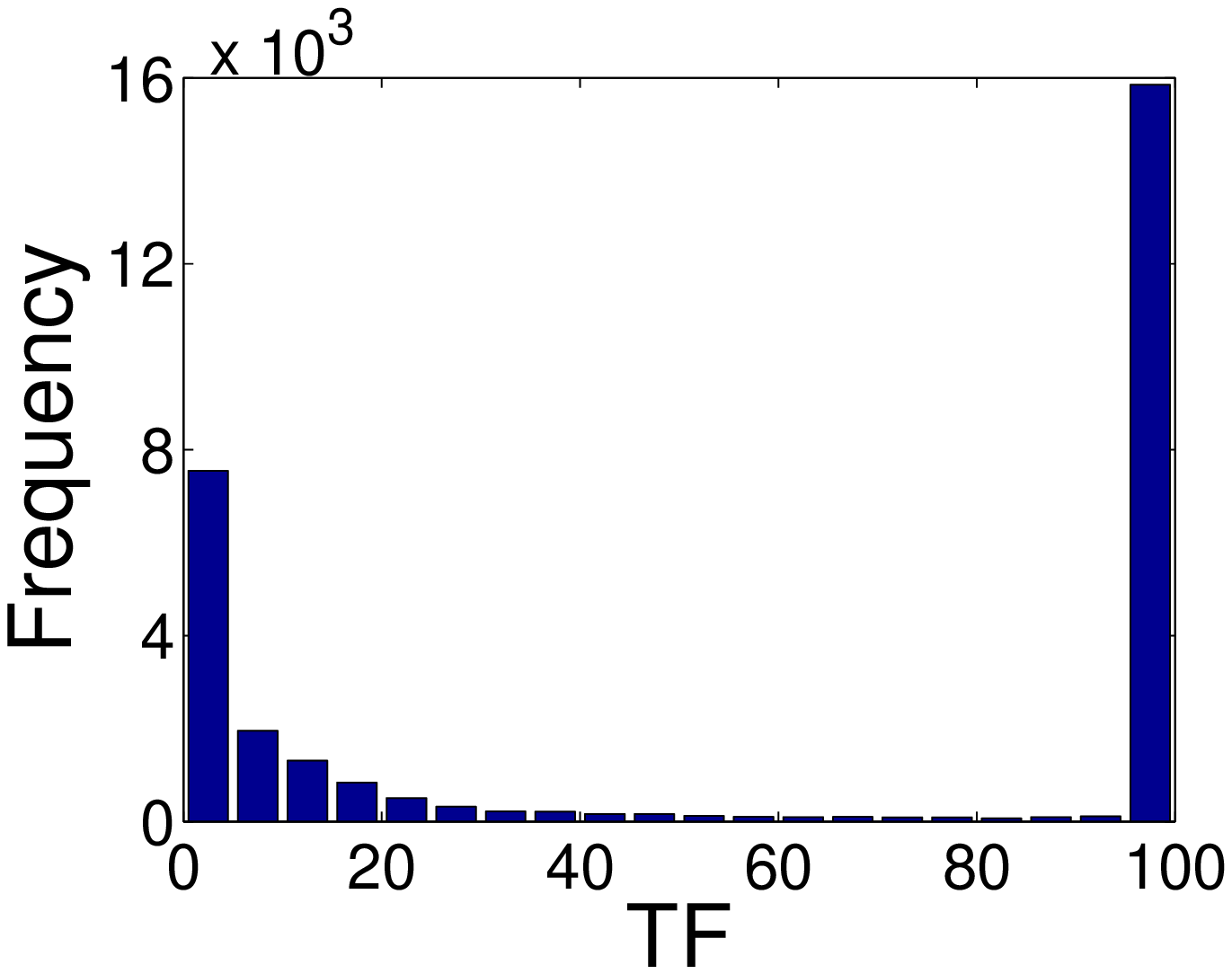}}
    \subfigure[$L=4$]{\label{fig:lostfload4}\includegraphics[scale=0.25]{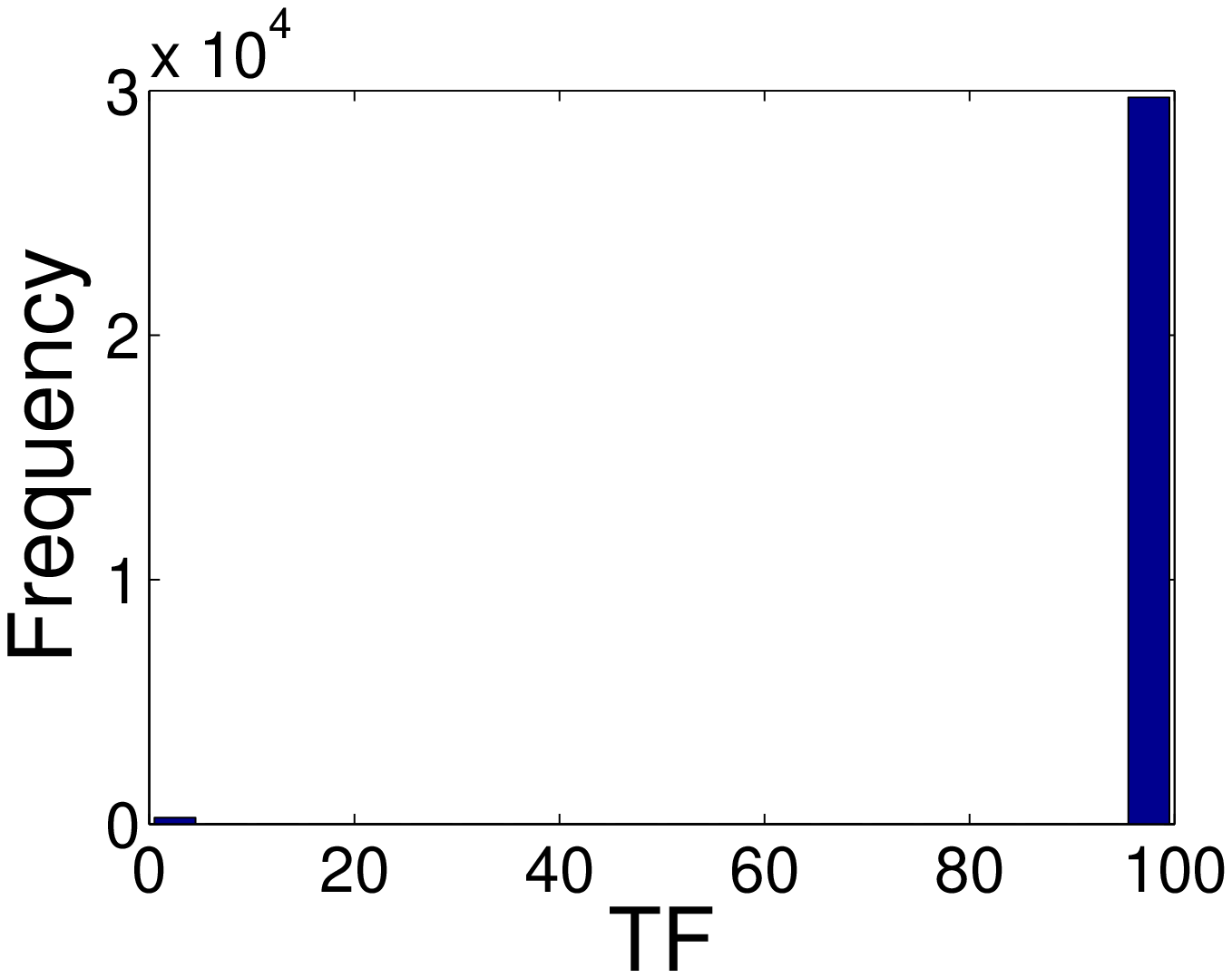}} \\
    \subfigure[$L=0.5$]{\label{fig:losttloadpt5}\includegraphics[scale=0.25]{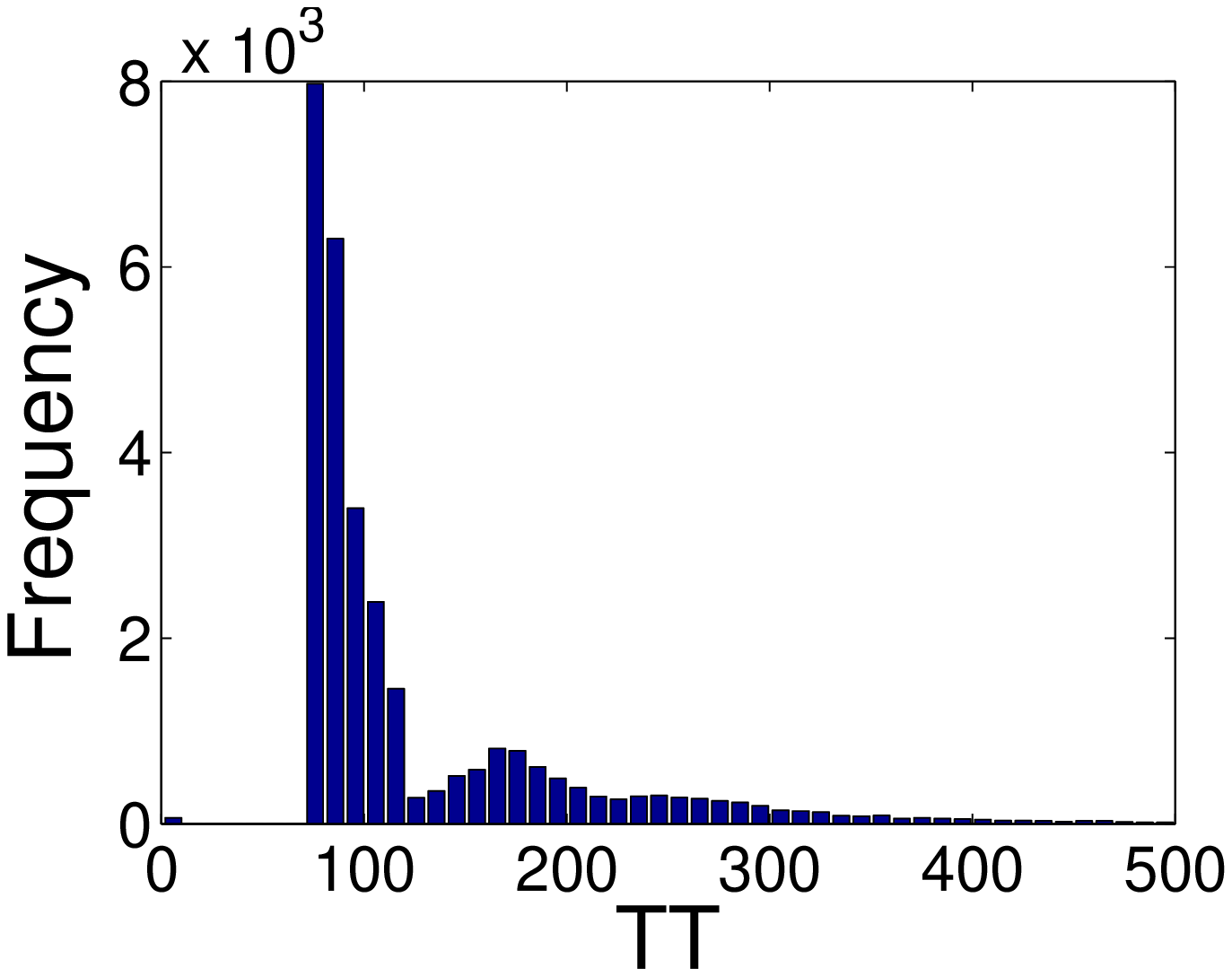}}
    \subfigure[$L=2$]{\label{fig:losttload2}\includegraphics[scale=0.25]{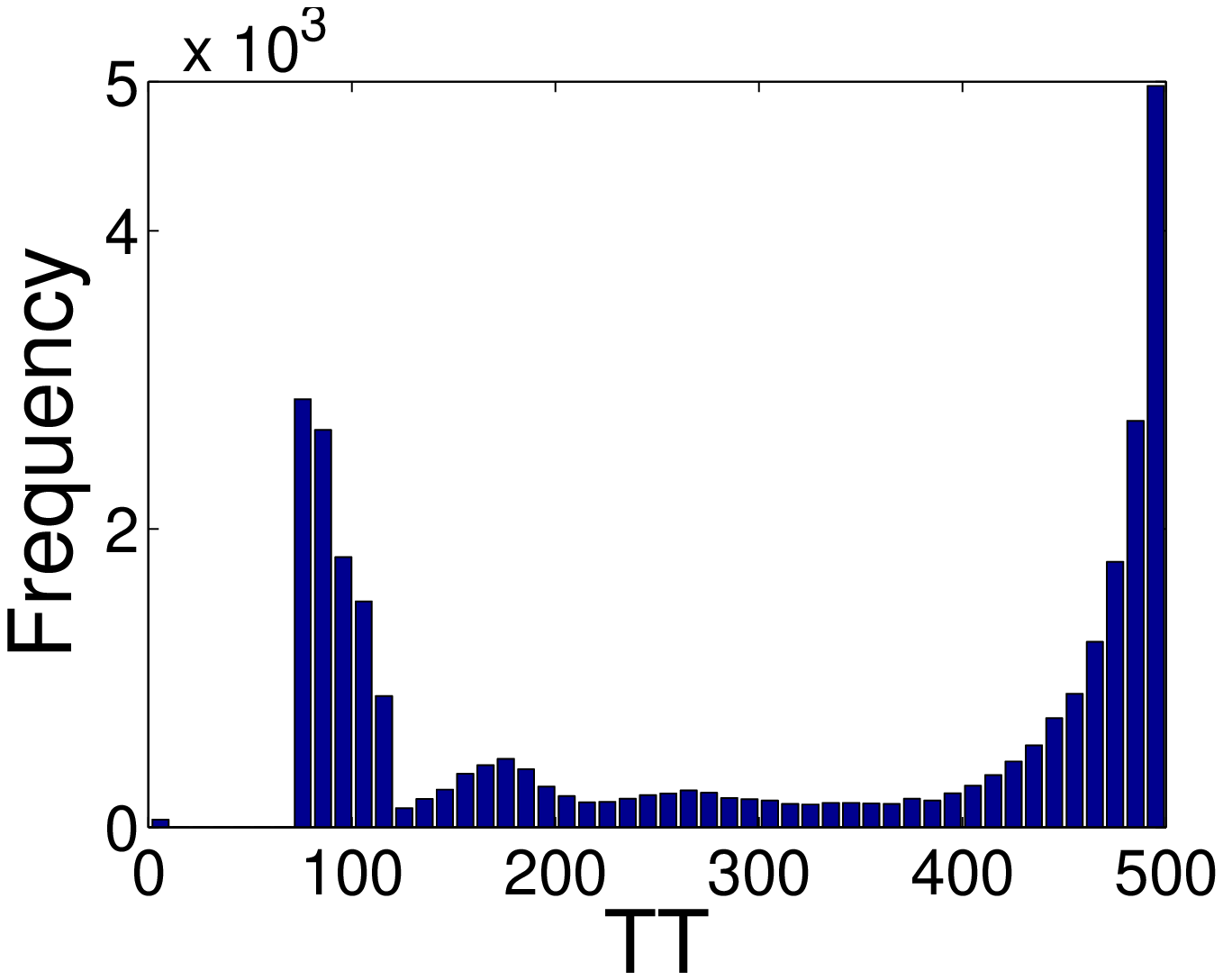}}
    \subfigure[$L=2.5$]{\label{fig:losttload2pt5}\includegraphics[scale=0.25]{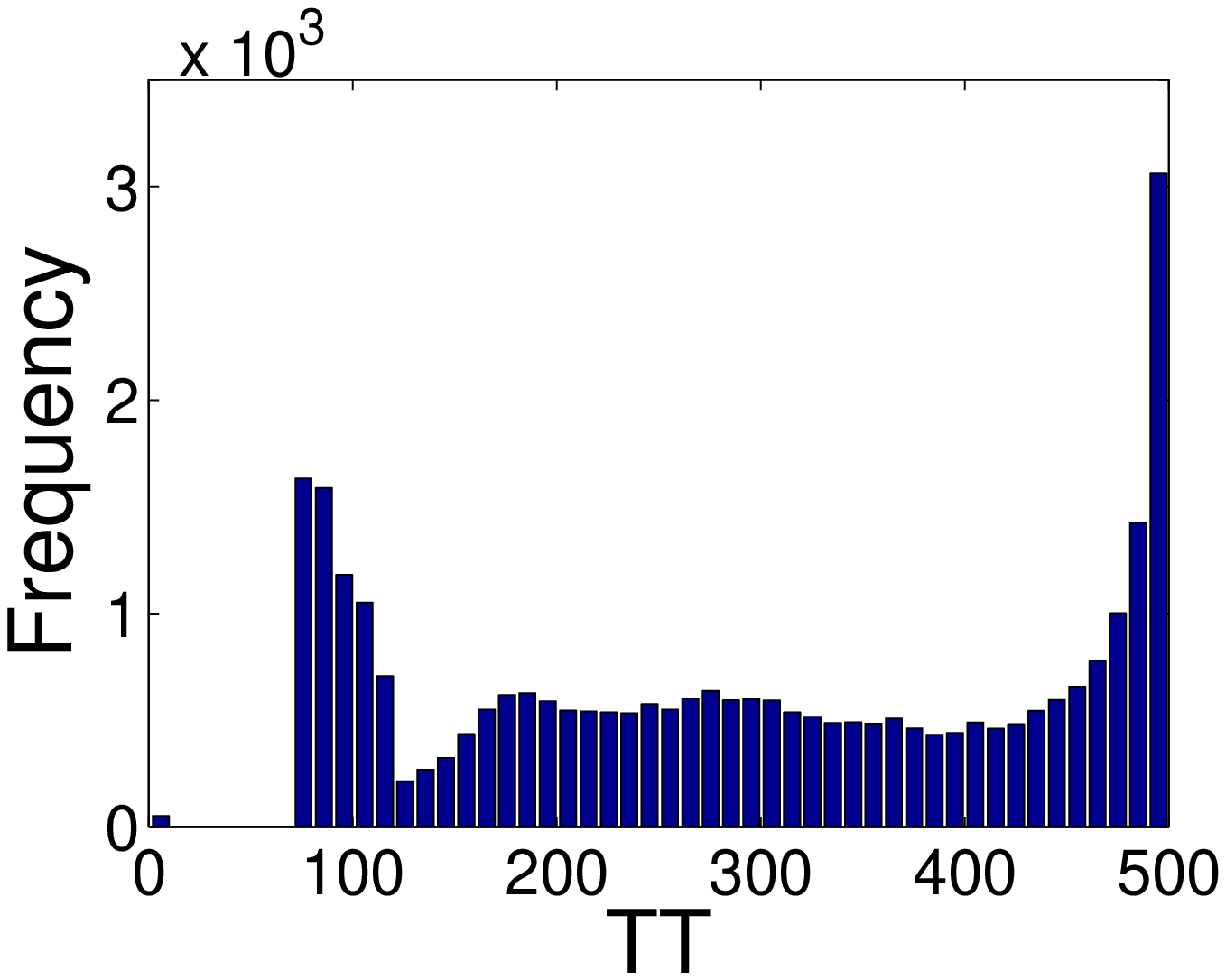}}
    \subfigure[$L=4$]{\label{fig:losttload4}\includegraphics[scale=0.25]{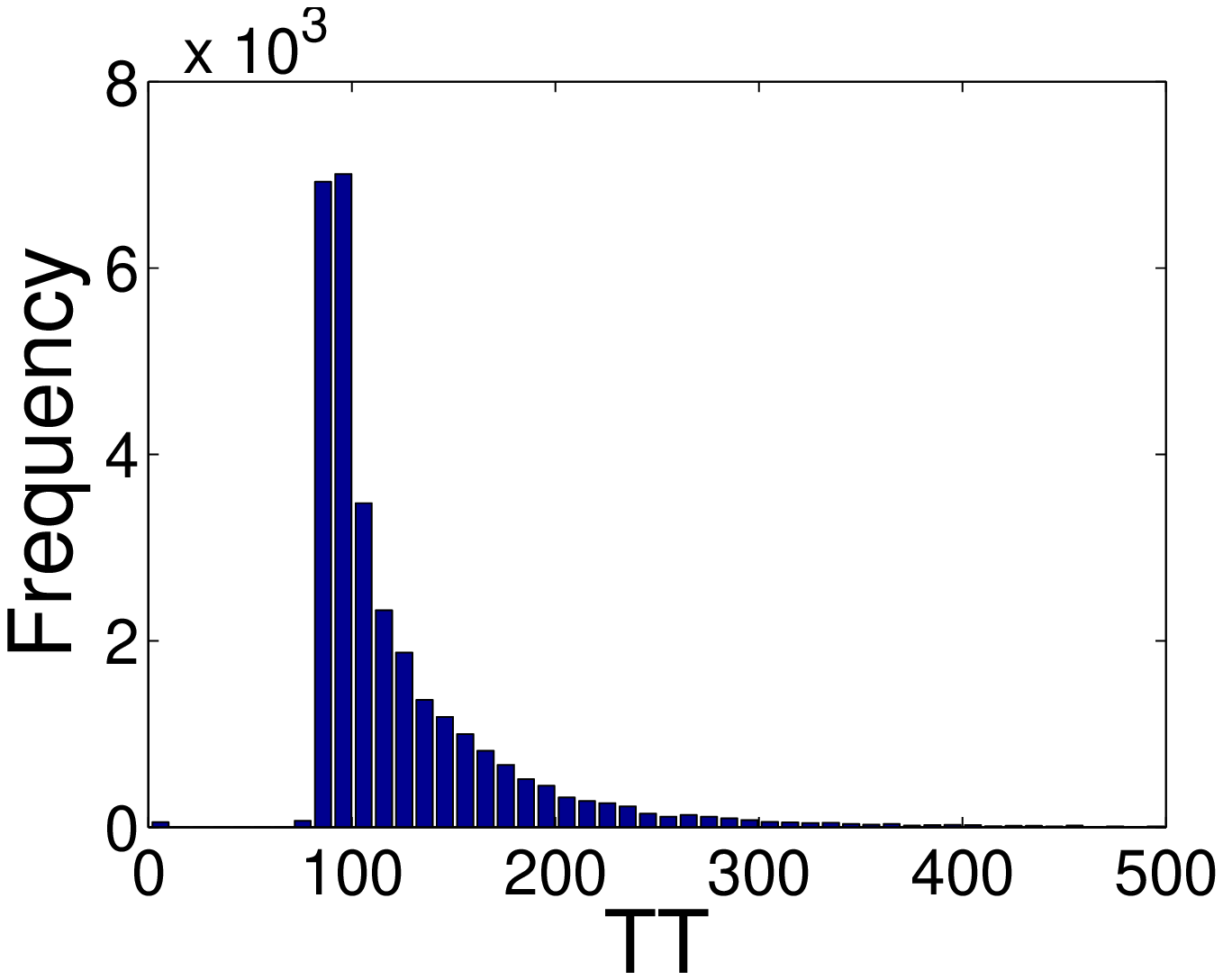}} \\
  \end{center}
  \caption{LOS model distributions. TF distributions (top row); TT distributions (bottom row).}
  \label{fig:tttfdist}
\end{figure*}

Before presenting our results we note a important difference between the CS and LOS models. In the LOS model, a component's load redistribution is the final step before it fails: once LOS dynamics is initiated the component will fail and an attempt will be made to redistribute its load. In the LOS model a component can undergo, at most, one load redistribution. In the CS model a component is essentially renewed through successful excess demand redistribution. The component fails only if the load redistribution is unsuccessful and the associated component queue $q_{ij}$ is overwhelmed. In the CS model a component can complete multiple excess demand redistributions and remain fully operational.

Another important point is regarding failed component load that is not successfully redistributed. In both models we shed the load and consider it lost. This is common in the context of internet traffic where packets are routinely discarded when routers are congested \cite{jacobson:1}. Similarly, power grid substations have mechanisms which take them offline during capacity/demand imbalances \footnote{M.J. Wald, R. Perez-Pena, N. Banerjee, \emph{The New York Times}, August 16, 2003}.

%\section{Results}

Next, we define two quantities measured in the simulations. During each simulation components fail as the system evolves in time. We denote by `TF' (Terminal Failure) the number of component failures at the end of a simulation. TF is a measure of the degree of system failure. We denote by `TT' (Terminal Time) the time step when the penultimate component failure occurred. TT can be interpreted as the time the system achieves a  pseudo steady-state.

The TF and TT distributions for different values of load initialization for the LOS model on a lattice configuration are shown in Fig.~\ref{fig:tttfdist}. At loading $L=0.5$ the system is far from critical. At these loading levels component failures are mainly due to components that were initialized to commence LOS and their subsequent load redistributions to weak components. At these loading levels the systems are resilient to chains of cascading failure triggered by load redistribution, this fact is indicated by Fig.~\ref{fig:lostfloadpt5}.

As the initialization load is increased, transition load conditions can be identified for $L=2, 2.5$. The bi-modal distributions in Fig.~\ref{fig:lostfload2},~\ref{fig:lostfload2pt5} resemble bathtub like curves that are commonly observed in reliability of complex systems \cite{meeker:1}. In Fig.~\ref{fig:lostfload2pt5} half the simulations represent systems with all components failing. The other half represent systems constituting partial component failures. This implies for approximately 50\% of the simulations the system strength topology is such that cascading chains of load redistributions are triggered which eventually bring down the entire system. For the other 50\% of simulations the system strength topology is strong enough to withstand the load redistributions thus avoiding a cascading chain of failures. Transition load settings are similar to `tipping points' or `critical thresholds' \cite{stanley:1, *marro:1}. In our simulation framework, at tipping-points systems may or may not, depending on system strength topology, descend into catastrophic failure.

%\section{Scaling phenomena}

\begin{figure}[!]
  \begin{center}
    \subfigure[Lattice network $L=0.5$]{\label{fig:lostttfscalept5}\includegraphics[scale=0.25]{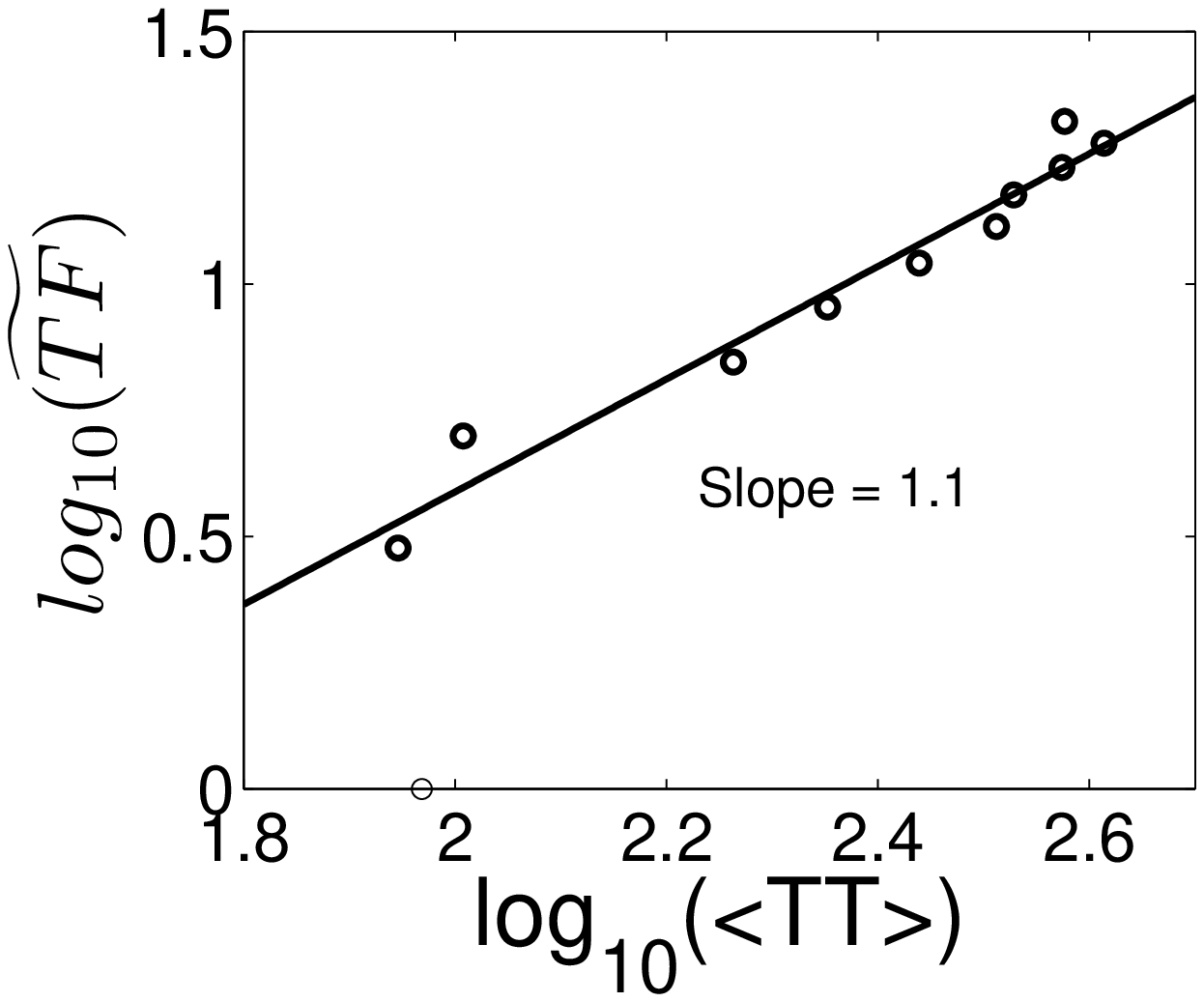}}
    \subfigure[Lattice network $L=1.5$]{\label{fig:lostttfscale1pt5}\includegraphics[scale=0.25]{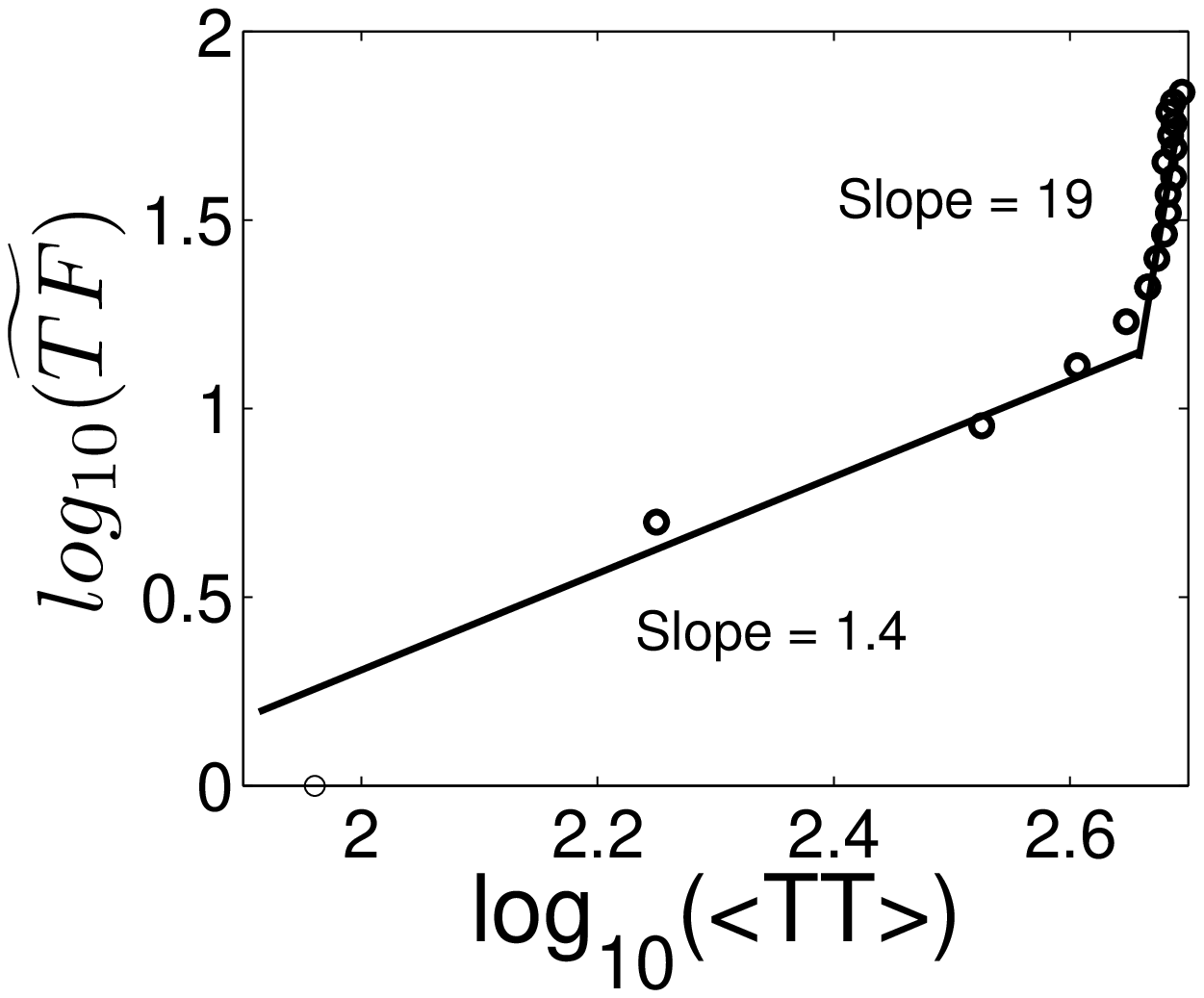}} \\
    \subfigure[Scale-Free network $L=0.5$, $\langle k\rangle=12$]{\label{fig:cstttfscale6}\includegraphics[scale=0.25]{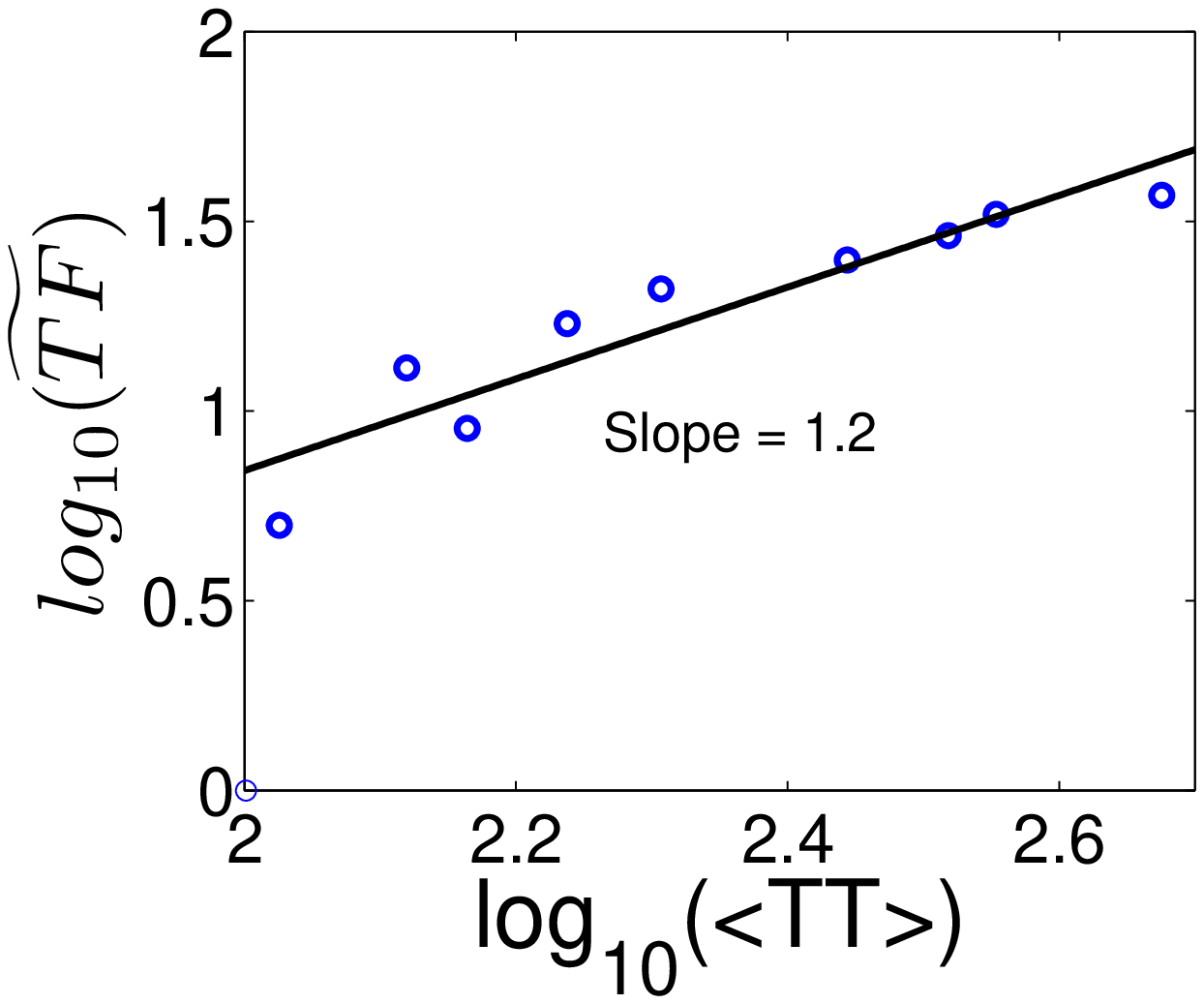}}
    \subfigure[Scale-Free network $L=1.5$, $\langle k\rangle=12$]{\label{fig:cstttfscale8}\includegraphics[scale=0.25]{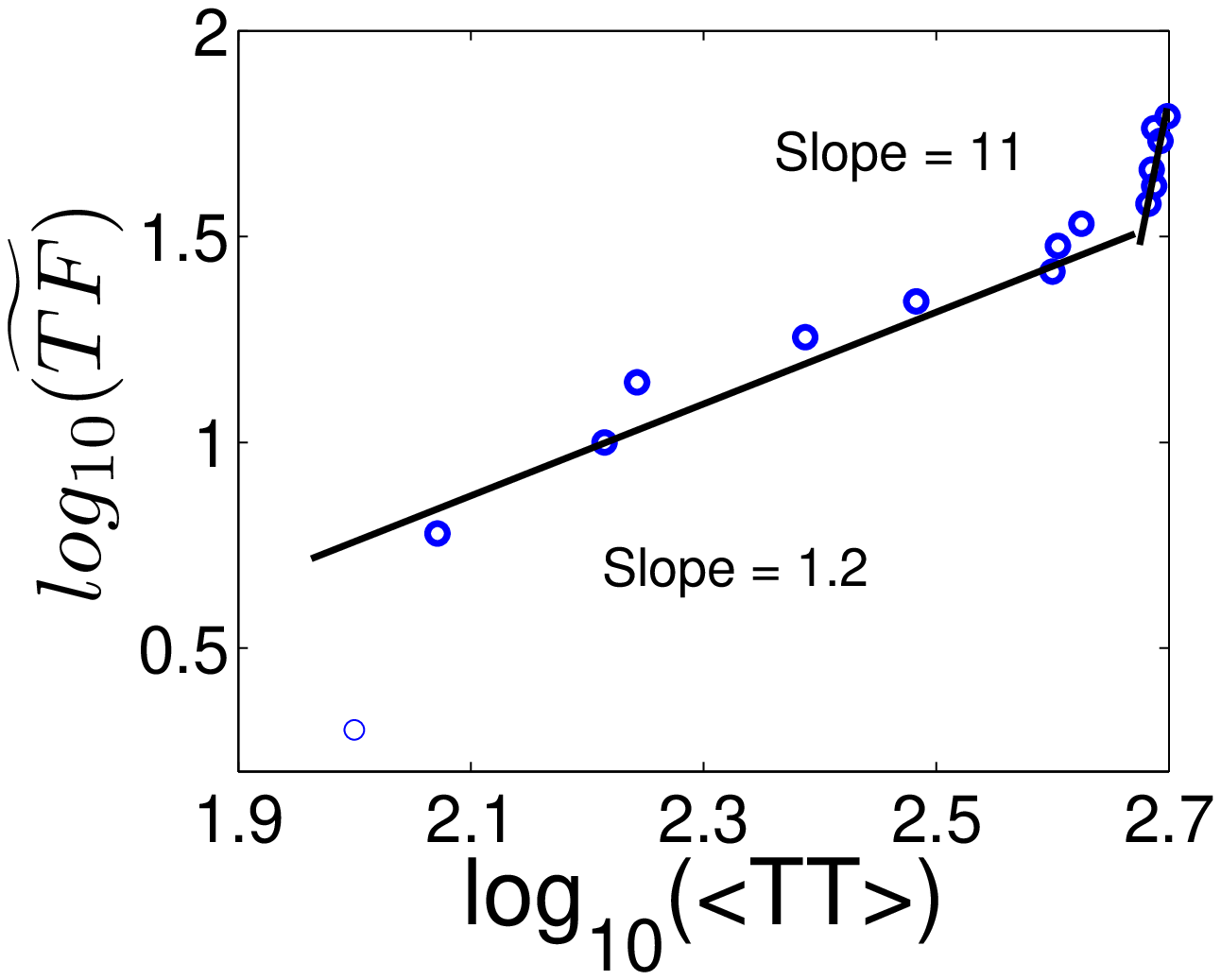}}
   \end{center}
  \caption{$\widetilde{TF}$ temporal scaling for the LOS model, lattice configuration (top row) and SF network (bottom row). The plots are on a base 10 log-log scale.}
  \label{fig:tttfscale}
\end{figure}

Recalling that TF represents the degree of system failure for a given simulation, let $\widetilde{TF}$ denote ``smoothed'' versions of TF. $\widetilde{TF}$ exhibits temporal scaling phenomena for the LOS model for load values lower than the transition load on both the lattice and SF networks of $\langle k\rangle=12$. $\widetilde{TF}$ is constructed in the following way. Referring to the TF and TT distributions in Fig.~\ref{fig:tttfdist}, each point in the TF distribution has a associated point in the TT distribution. For a particular system loading $L$, we bin the TF distribution in groups of 4 in ascending order and denote them by $\widetilde{TF}_i$ with value set to the minimum TF value in the bin \footnote{The results still hold if we set the value to the mean or max of the bin. The objective of binning is to smooth or denoise the data}. For each bin we find the corresponding values in the TT distribution and compute their mean, denoted $\langle TT \rangle_{i}$. We illustrate the temporal scaling phenomena of $\widetilde{TF}$ in Fig.~\ref{fig:tttfscale} for different values of loading for the LOS model in both lattice and SF network configurations. In Fig.~\ref{fig:tttfscale}, $\widetilde{TF}$ versus $\langle TT \rangle$ is plotted in a log-log scale. Each circle in the figures represents the mean of a TT distribution conditioned on a $\widetilde{TF}_{i}$.

From these figures the following scaling relation is established for loading values far below the critical load,

\begin{equation}
\langle TT \rangle = \kappa \widetilde{TF}^{\tau}
\label{eq:tttf}
\end{equation}

Table~\ref{tab:ttfscale} tabulates the numerical values for $\kappa$ and $\tau$ for different values of load.  At low values of load, Figures~\ref{fig:lostttfscalept5} and~\ref{fig:cstttfscale6}, the logarithm of $\widetilde{TF}$ scales linearly versus the logarithm of $\langle TT \rangle$. As the initial load setting is increased, a breakpoint develops and the $\widetilde{TF}$'s separate into two different log-log linear scales, as illustrated in Figs.~\ref{fig:lostttfscale1pt5},~\ref{fig:cstttfscale8}. The slopes of the figures indicate the second group of $\widetilde{TF}$'s have faster transition dynamics to $\langle TT \rangle$ compared to the first $\widetilde{TF}$ group. Table~\ref{tab:ttfscale} also tabulates the break point $\widetilde{TF}$ when the switch to faster transition dynamics occurs and the residual error of the data fit.

\begin{figure}[!]
  \begin{center}
    \subfigure[First order transition at $L=4$]{\label{fig:losfirstorder}\includegraphics[scale=0.25]{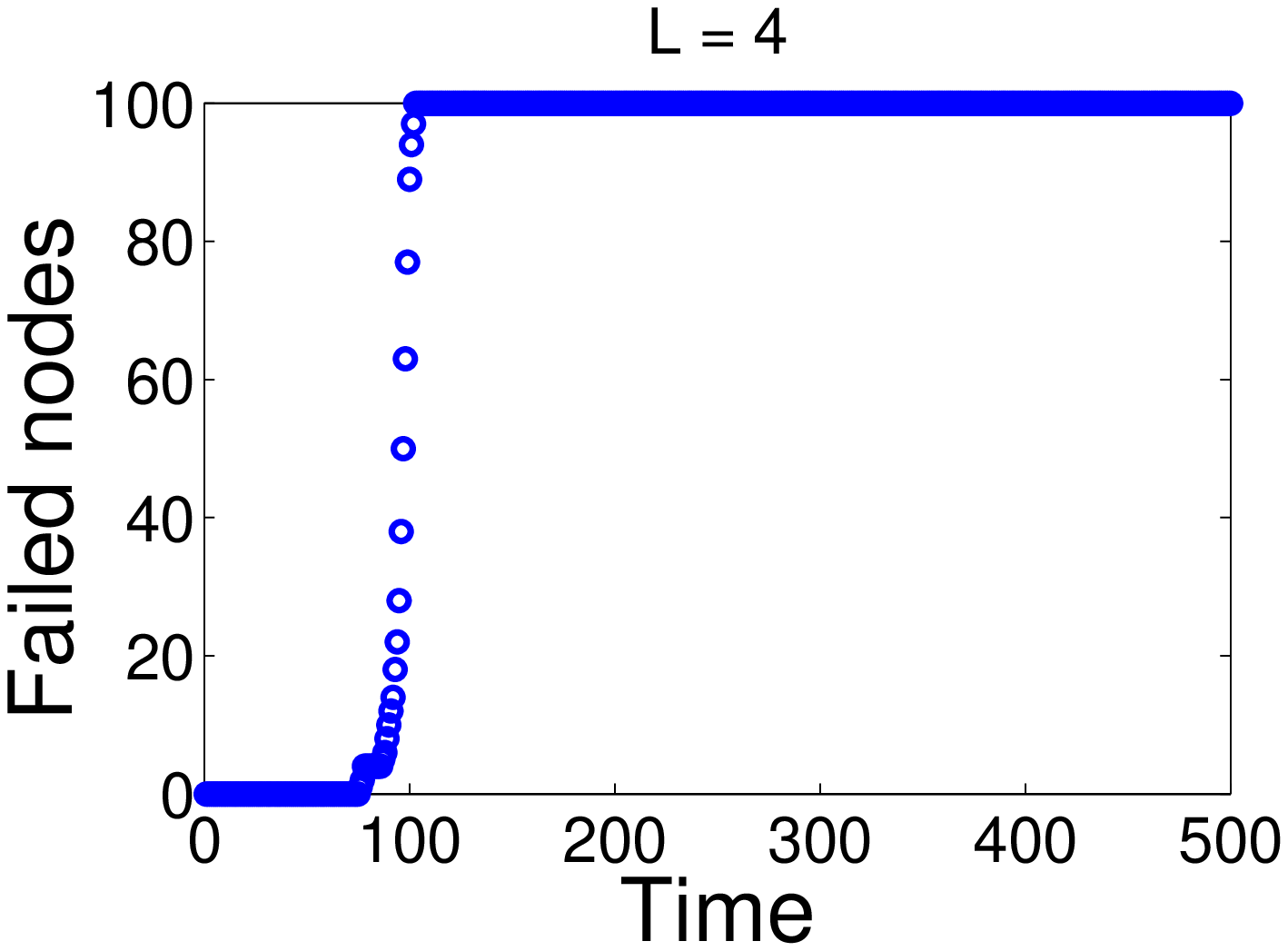}}
    \subfigure[Second order transition at $L=2$]{\label{fig:lossecondorder}\includegraphics[scale=0.25]{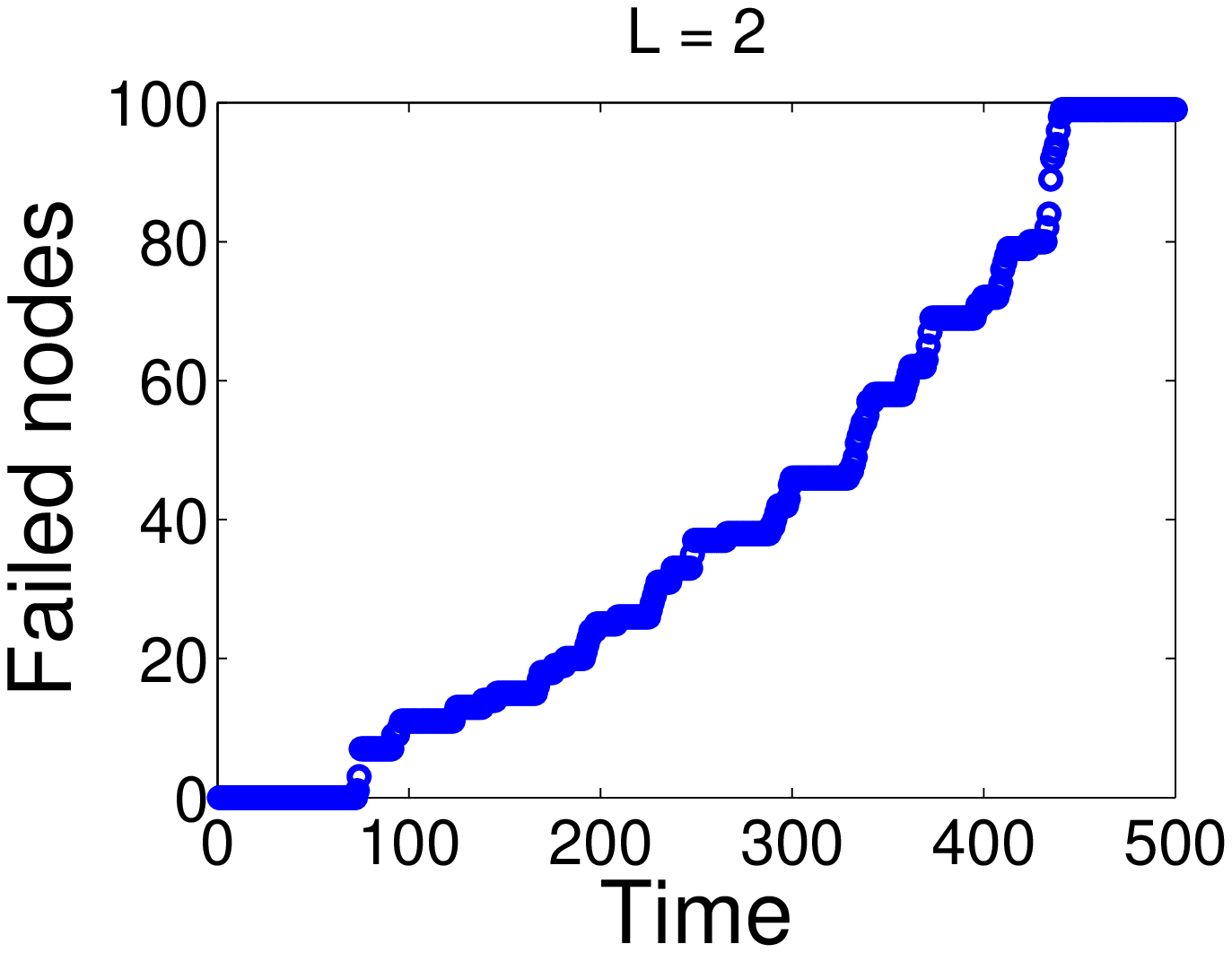}}
  \end{center}
  \caption{Phase transition of the LOS lattice model}
  \label{fig:phasetransition}
\end{figure}

All systems in the LOS lattice configuration undergo complete failure at the critical load $L=4$, as seen in Fig.~\ref{fig:lostfload4} and $\widetilde{TF}$ is characterized by a first-order phase transition into the critical state [Fig.~\ref{fig:losfirstorder}]. We define the critical load as the load required for complete system failure, TF = 99 or 100, with probability greater than 0.95. For the LOS model on SF networks the critical load is slightly higher at $L=4.5$ for $\langle k \rangle =12$, $L=5$ for $\langle k \rangle =8$ and $L=5.5$ for $\langle k \rangle =4$. Here we recall that component strengths are initialized in the real number interval $\mathcal{U}[6,10]$ with $\eta=0.7$, meaning that for $L > 6 \times 0.7 = 4.2$ a considerable number of components will be initialized in the stressed mode $L_{ij} > \eta S_{ij}$. To induce system failure in LOS SF networks, with decreasing $\langle k \rangle$ more and more components need to be initialized to commence failure dynamics. The implication being that the LOS model is increasingly resilient to system failure with decreasing average network connectivity $\langle k \rangle$. This result is in agreement with \cite{albert:1} which demonstrate that scale free networks are more resilient to random errors or failure \footnote{In the literature there is a distinction between errors and attacks. Attacks target specific nodes where as random nodes are susceptible to errors. In this work we commence LOS dynamics for random components} compared to other network topologies.

%\section{Critical behavior}

\begin{figure}[!]
  \begin{center}
    \subfigure[$L=4$ LOS Lattice model, Fit to $P(TT)=\alpha TT^{-\beta}$]{\label{fig:losttlogload4}\includegraphics[scale=0.25]{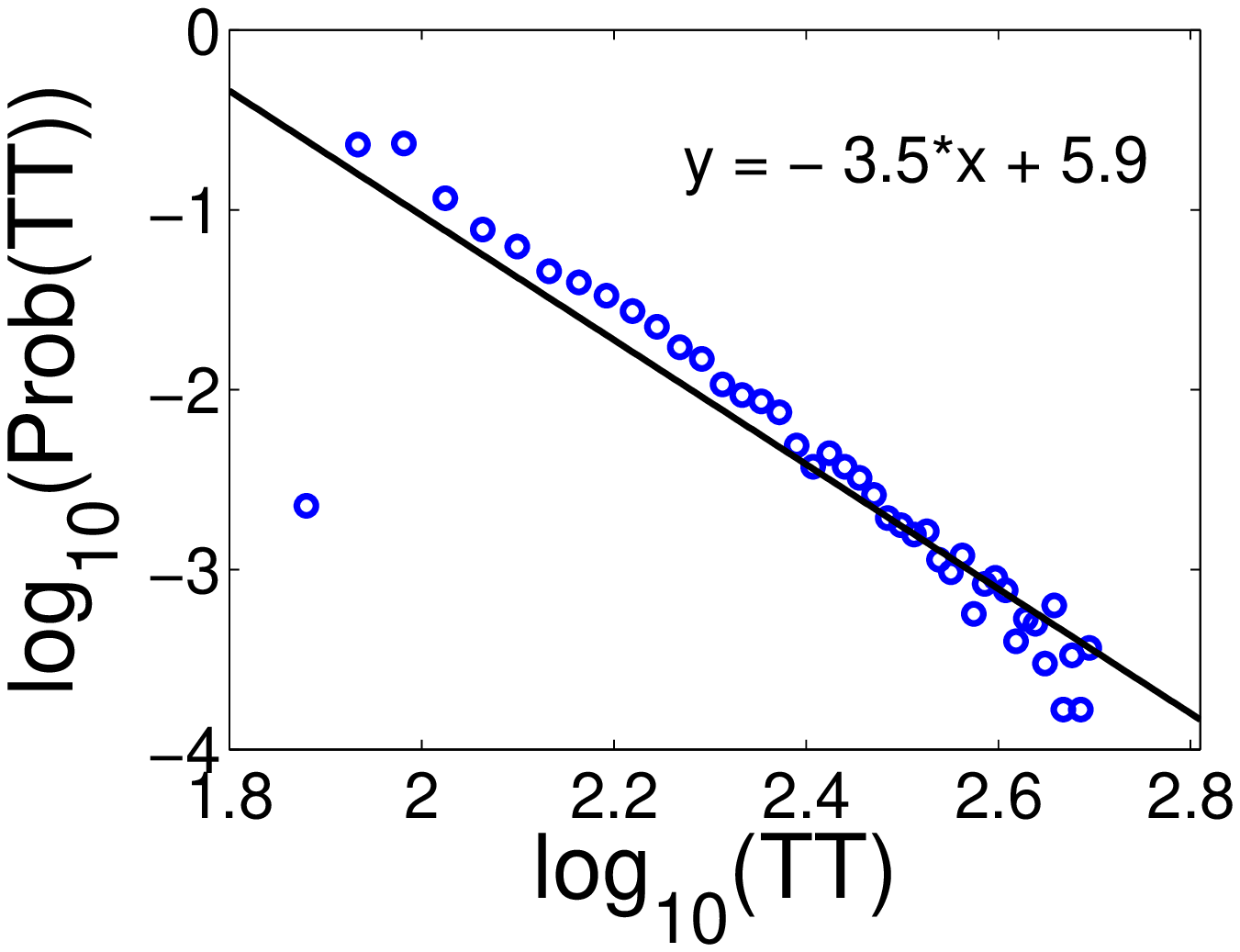}}
    \subfigure[$L=4.5$, $\langle k\rangle=12$ LOS SF model, Fit to $P(TT)=\alpha TT^{-\beta}$]{\label{fig:lossfttlogload4pt5}\includegraphics[scale=0.25]{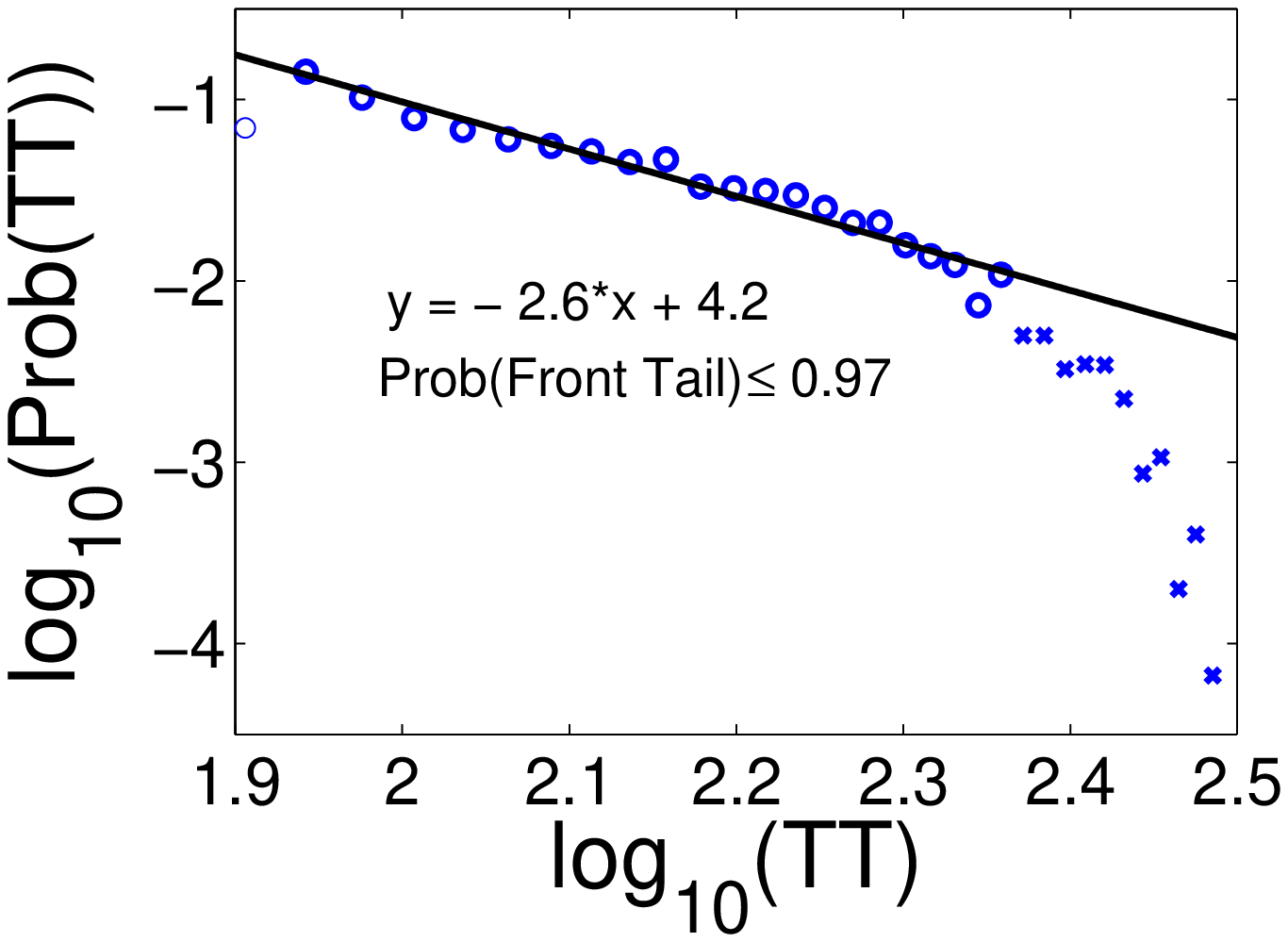}} \\
    \subfigure[$L=5$, $\langle k\rangle=8$ LOS SF model]{\label{fig:lossfttlogload5}\includegraphics[scale=0.25]{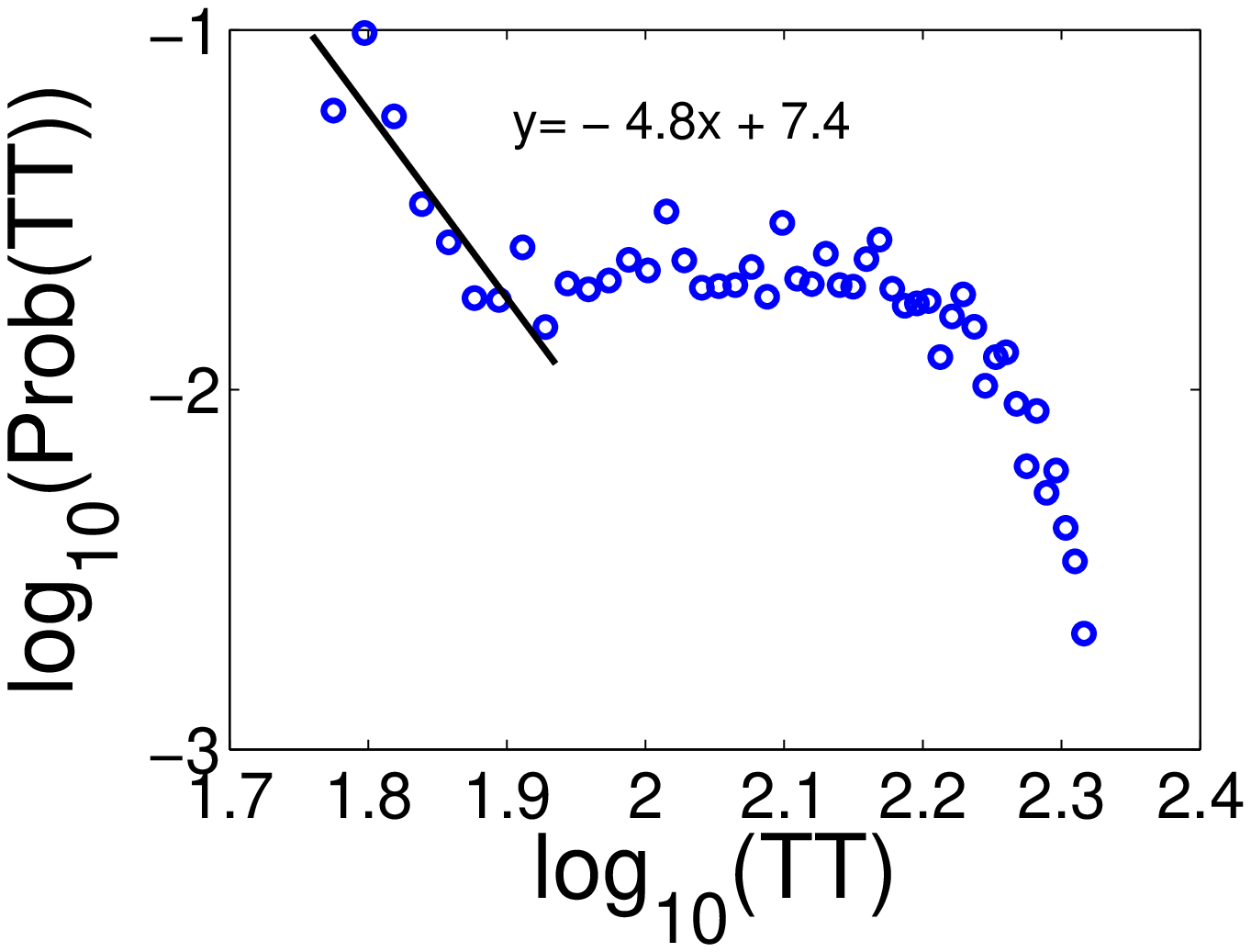}}
    \subfigure[$L=5.5$, $\langle k\rangle=4$ LOS SF model]{\label{fig:lossfttlogload5pt5}\includegraphics[scale=0.25]{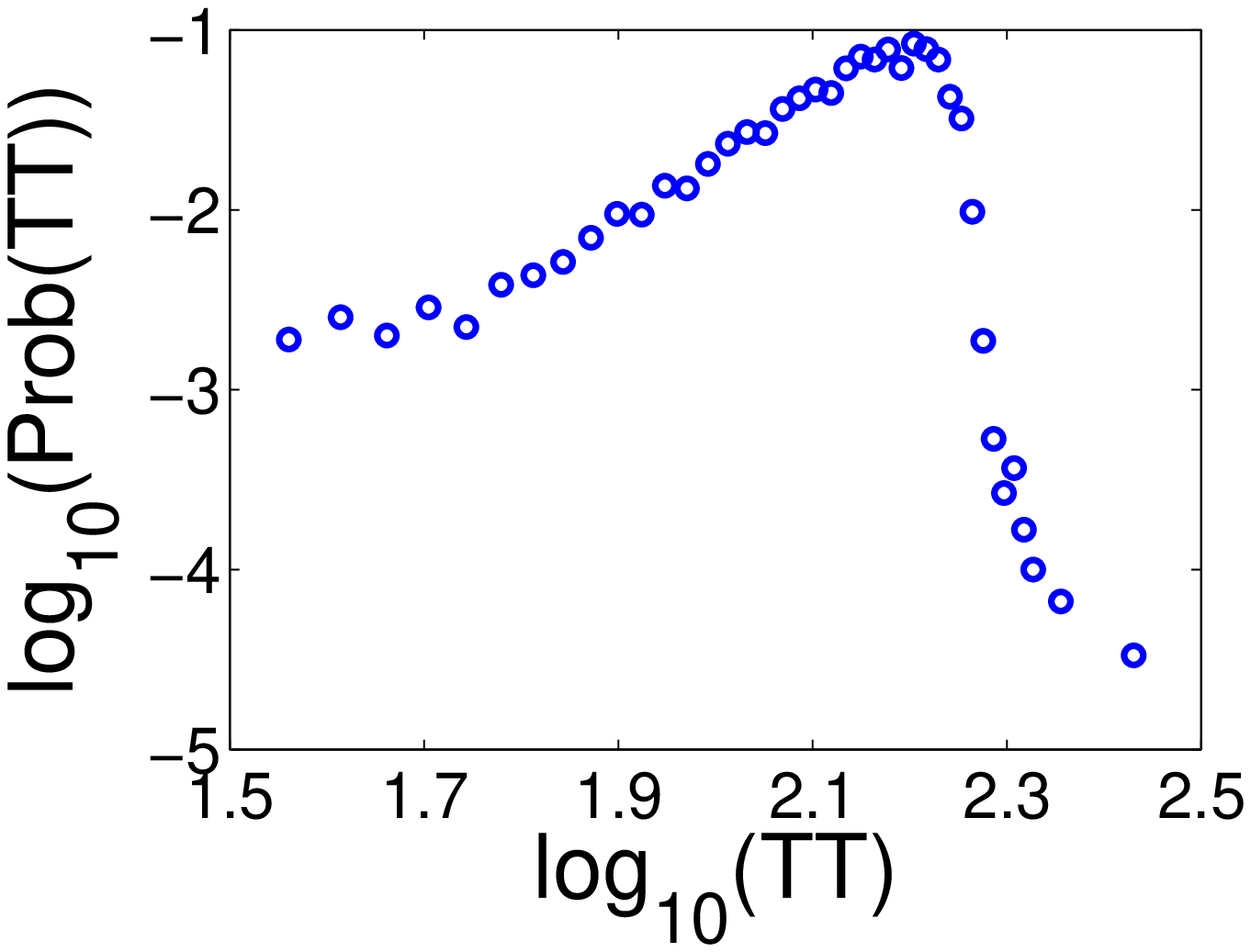}}
    \end{center}
  \caption{LOS model TT distributions at the critical load. Note the breakdown in power law scaling with decreasing average network connectivity.}
  \label{fig:ttcomparelos}
\end{figure}

The TT distribution fit for the LOS model at the critical loads is shown in Fig.~\ref{fig:ttcomparelos}. The LOS model on a lattice configuration for $L=4$ is shown in Fig.~\ref{fig:losttlogload4}. In the lattice configuration the LOS model fits a power law distribution $Prob(TT) = \alpha TT^{-\beta}$ with $\beta=3.5$ and $\alpha=10^{5.9}$. The LOS model on SF networks with $L=4.5$, $\langle k \rangle = 12$ is shown in Fig~\ref{fig:lossfttlogload4pt5}. With probability 0.97 the model fits a power law distribution $Prob(TT) = \alpha TT^{-\beta}$ with $\beta=2.6$ and $\alpha=10^{4.2}$. The LOS model on SF networks with $L=5$, $\langle k \rangle = 8$ is shown in Fig~\ref{fig:lossfttlogload5} and $L=5.5$, $\langle k \rangle = 4$ is shown in Fig~\ref{fig:lossfttlogload5pt5}. As can be seen from the figures, at the critical load as average degree $\langle k \rangle$ decreases the TT distributions loose their power-law scaling. Implying at the critical load, the LOS model looses power-law scale invariance in system failure time distribution with decreasing average network connectivity $\langle k \rangle$. %This indicates the failure time distributions for SF networks for $\langle k \rangle = 2, 4$ have multiscaling failure mechanisms \cite{lvov:1}.

%\section{Phase diagram}

The phase diagram of the LOS lattice model is shown in Fig.~\ref{fig:phasetransition}. The LOS lattice model demonstrates phase diagrams similar to both first-order and second-order phase transitions. At the critical load $L=4$, cascades of load redistributions induce massive failure causing all systems to fail as shown in Fig.~\ref{fig:lostfload4}. The corresponding first-order phase diagram is shown in Fig.~\ref{fig:losfirstorder}. The loading at $L=4$ is such that cascading load redistributions induce failure with minimal LOS dynamics. At transition loadings $L=2$, systems undergoing complete failure [refer to Figs.~\ref{fig:lostfload2},~\ref{fig:lostfload2pt5}] exhibit second-order phase transitions as shown in Fig.~\ref{fig:lossecondorder}. For second-order phase transitions, the transition to complete system failure is a gradual process involving repetition of LOS dynamics and load redistributions cascading from one component to the next.

Here we note, first and second-order phase transitions for traffic congestion in complex networks have also been reported in \cite{martino:1,echenique:1}. In \cite{martino:1}, the authors show that by increasing the probability of node congestion (from $\bar{\eta}=0.05$ to $\bar{\eta}=0.95$, where $\bar{\eta}$ is a parameter to control node congestion probability) the traffic flow phase diagram switches from second-order to first-order. On the other hand in \cite{echenique:1}, the first or second-order phase transitions depend on the particular traffic routing protocol utilized (shortest-path routing versus traffic-aware routing). In comparison for the LOS model, by increasing the load from $L=2$ to $L=4$, the component failure phase diagram switches from second-order to first-order.

%\section{Failure modes and extremal behavior}

\begin{figure*}[!]
  \begin{center}
    \subfigure[$\lambda=5$]{\label{fig:cstflambda5}\includegraphics[scale=0.25]{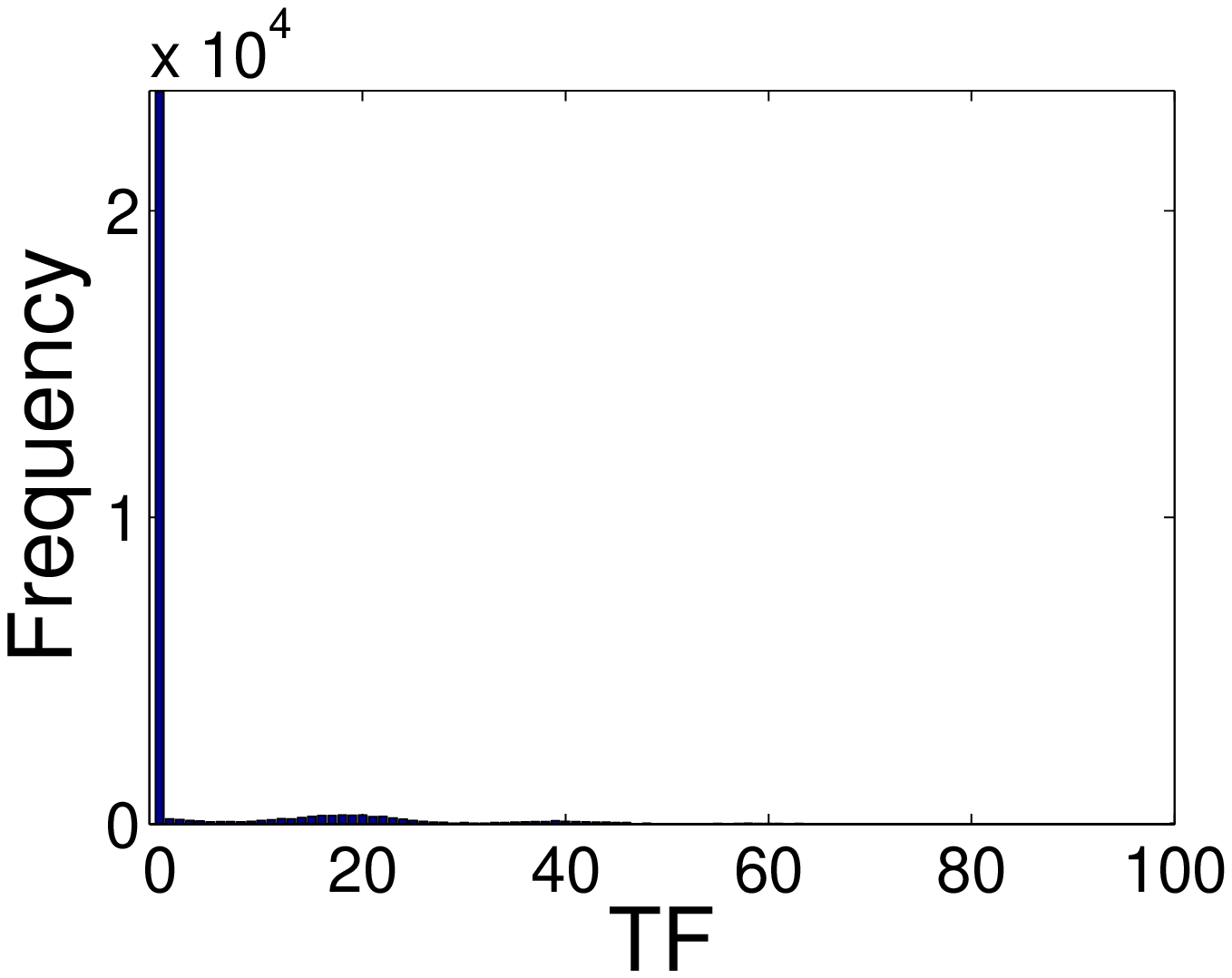}}
    \subfigure[$\lambda=6$]{\label{fig:cstflambda6}\includegraphics[scale=0.25]{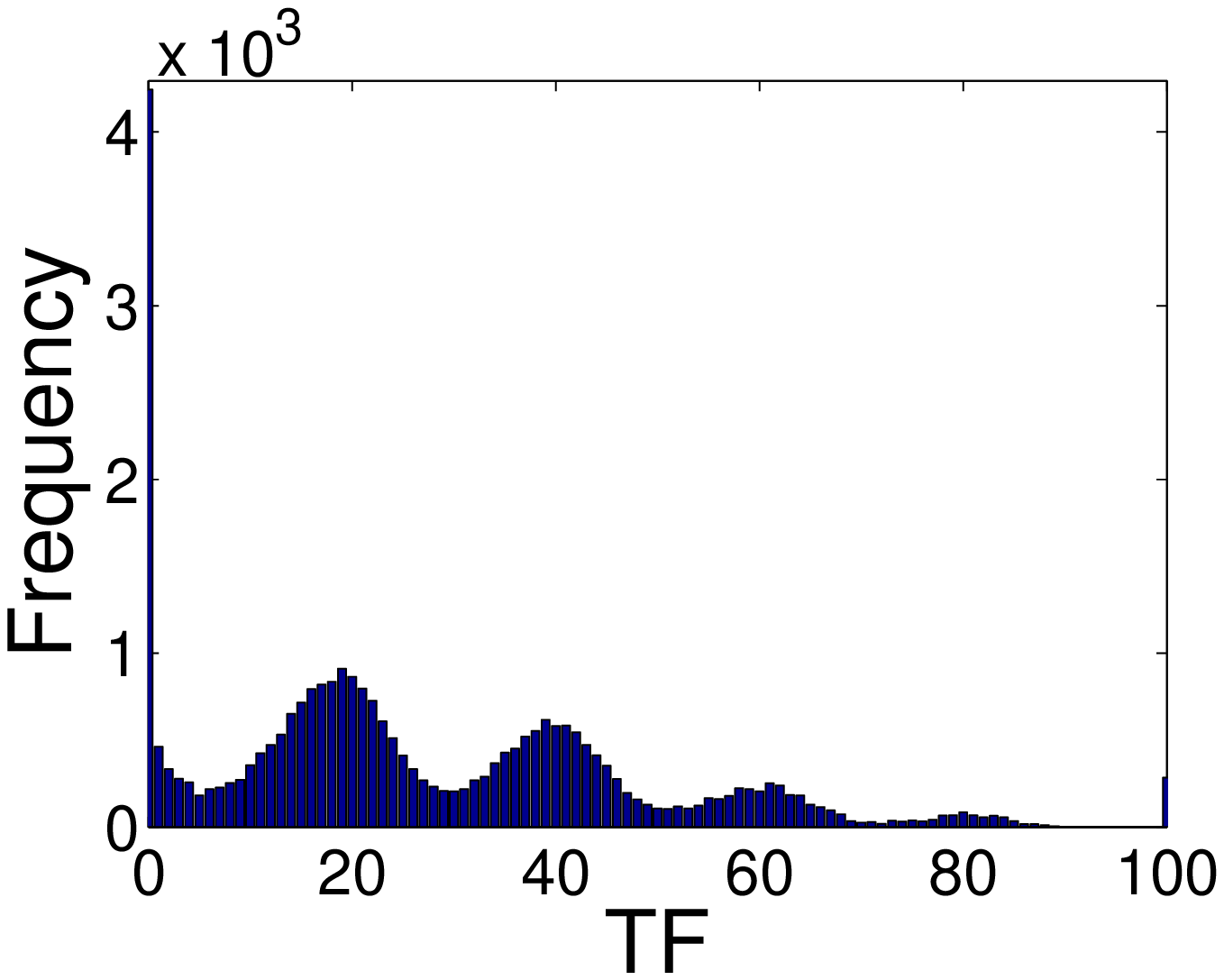}}
    \subfigure[$\lambda=7$]{\label{fig:cstflambda7}\includegraphics[scale=0.25]{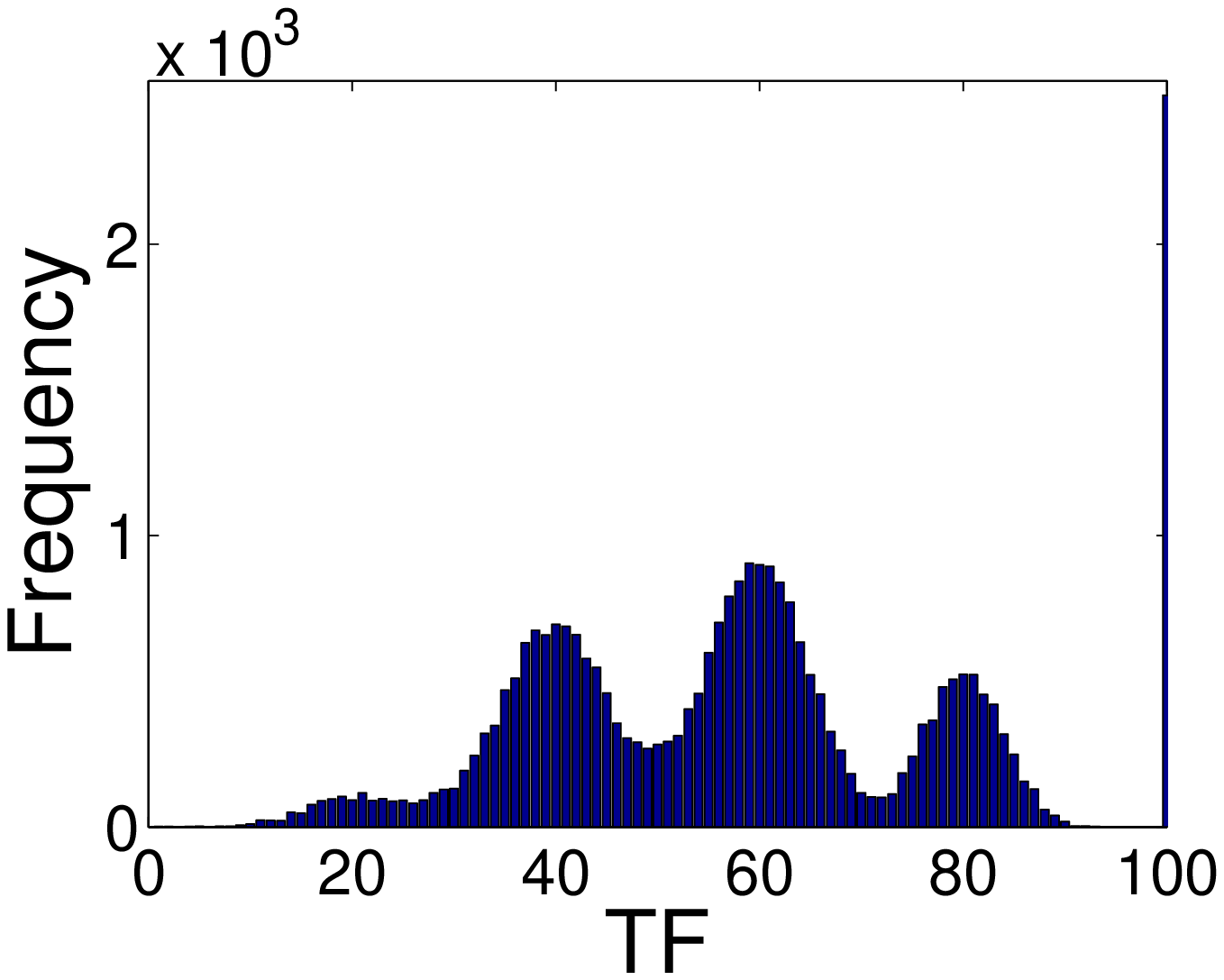}}
    \subfigure[$\lambda=9$]{\label{fig:cstflambda9}\includegraphics[scale=0.25]{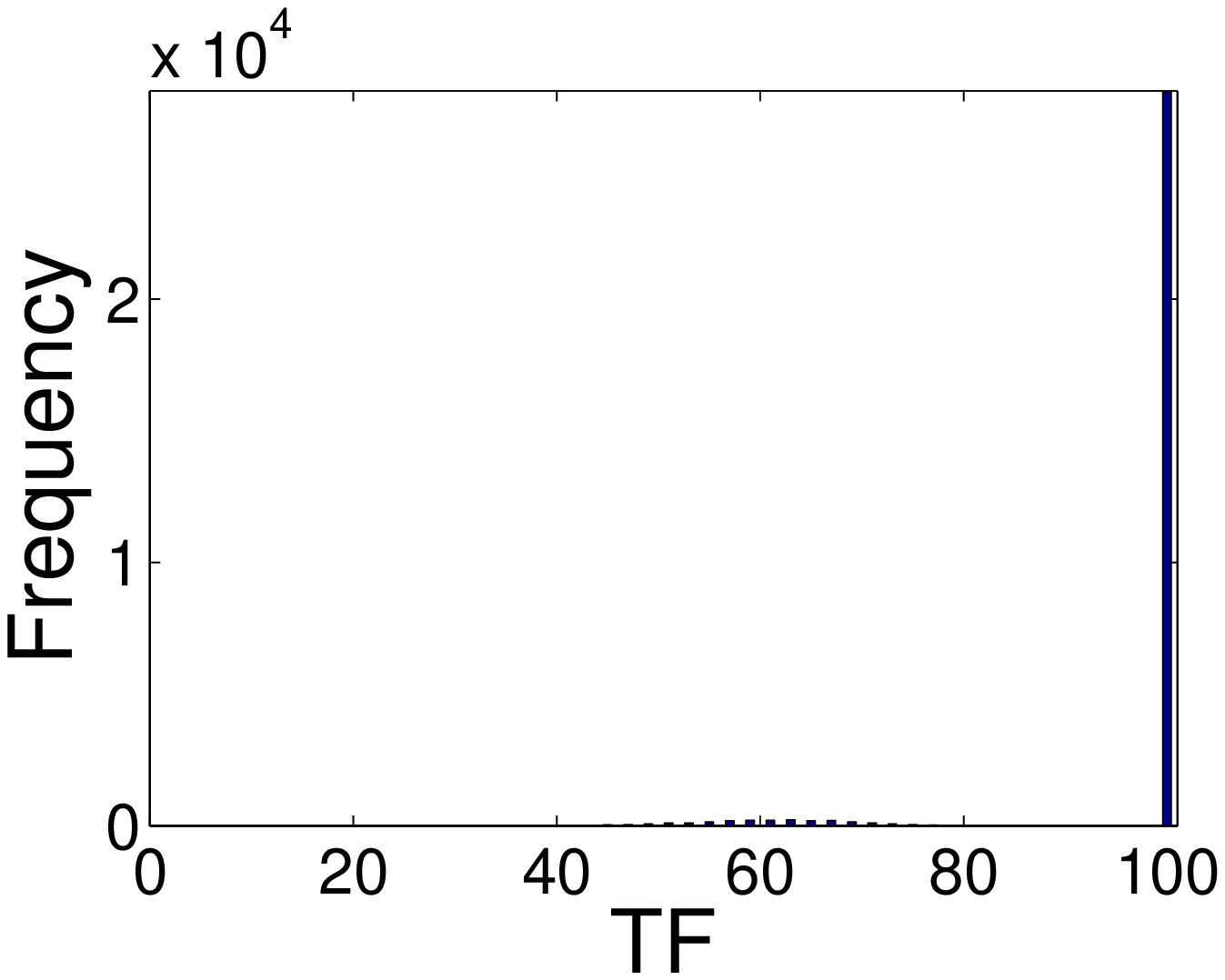}} \\
    \subfigure[$\lambda=5$]{\label{fig:csttlambda5}\includegraphics[scale=0.25]{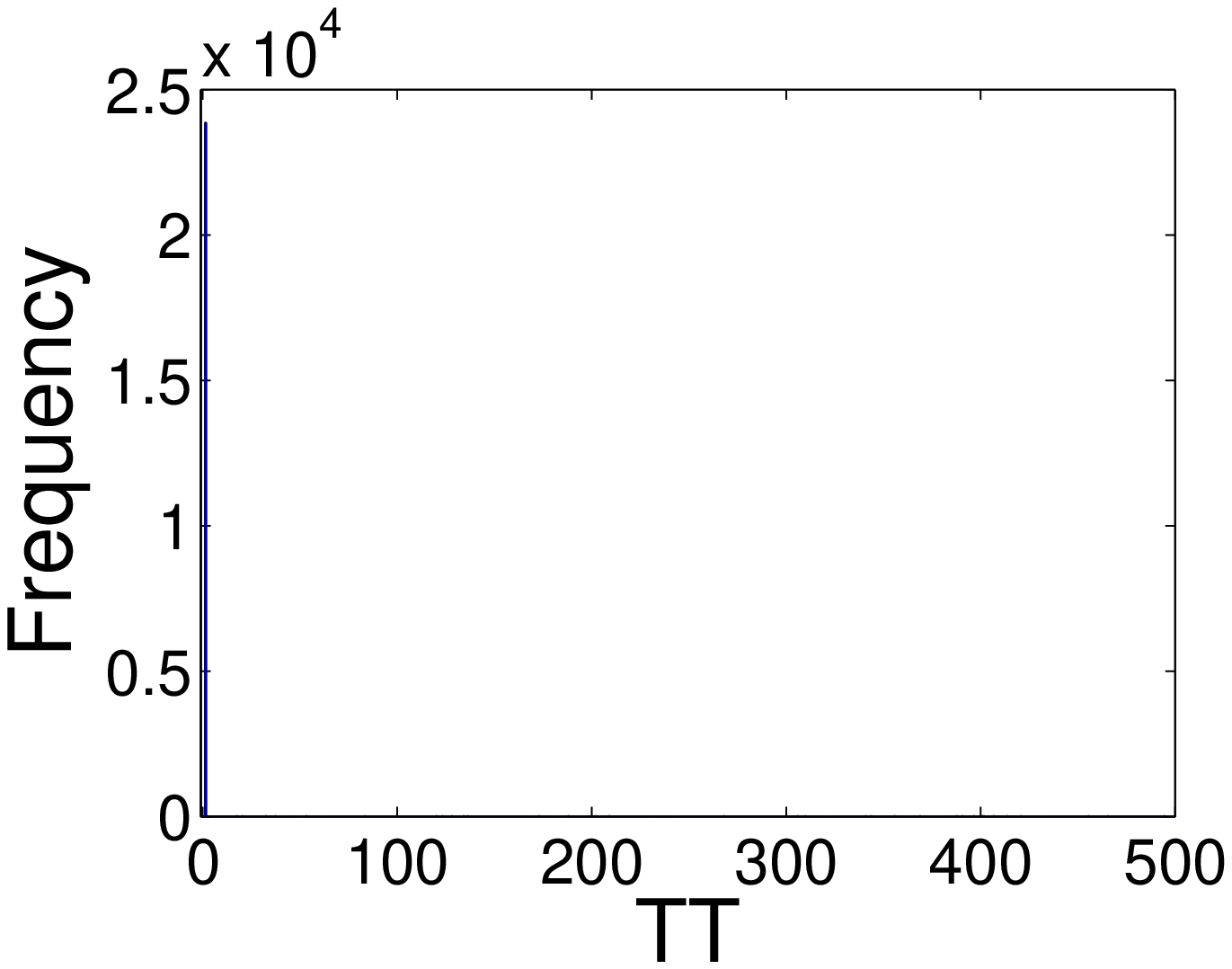}}
    \subfigure[$\lambda=6$]{\label{fig:csttlambda6}\includegraphics[scale=0.25]{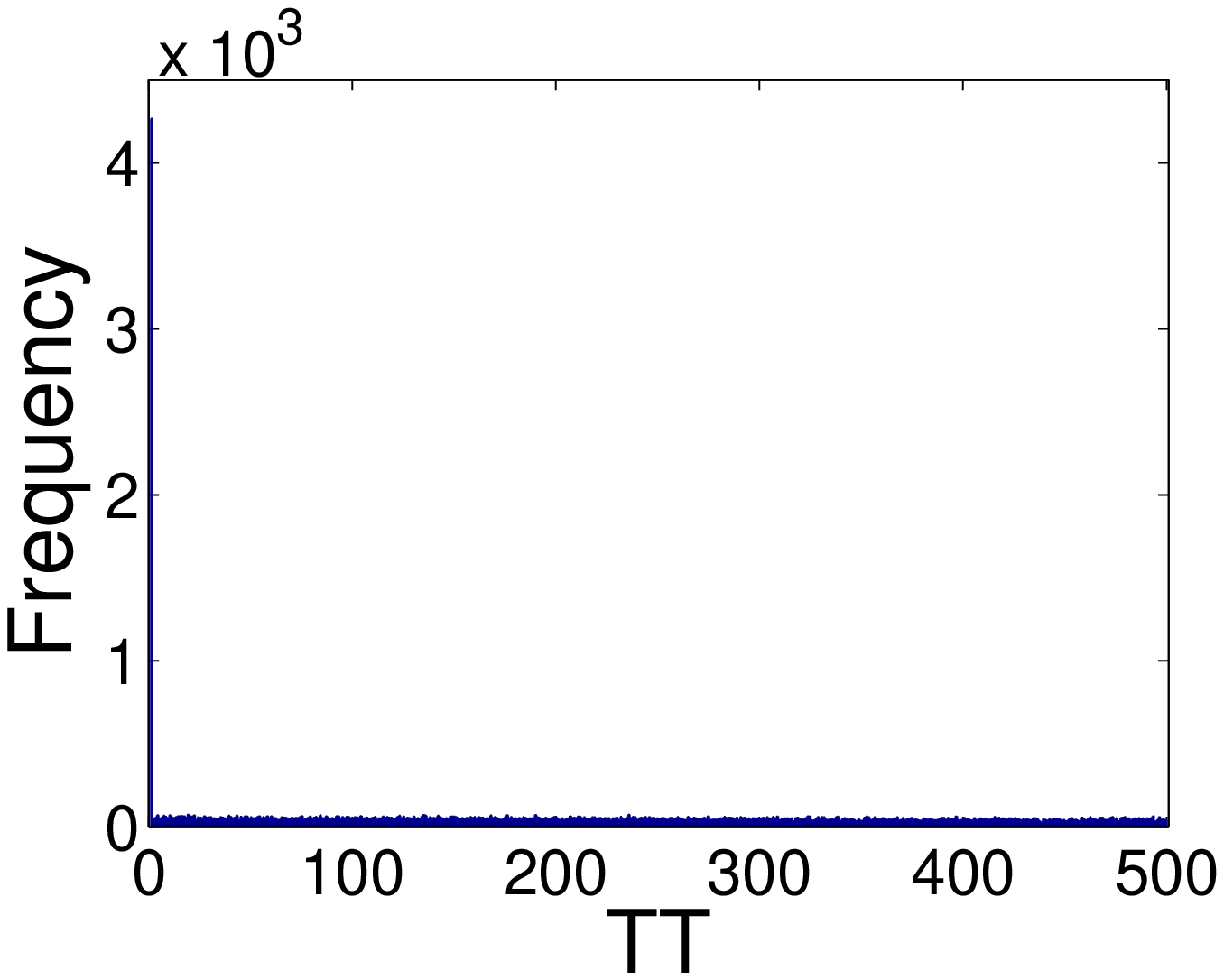}}
    \subfigure[$\lambda=7$]{\label{fig:csttlambda7}\includegraphics[scale=0.25]{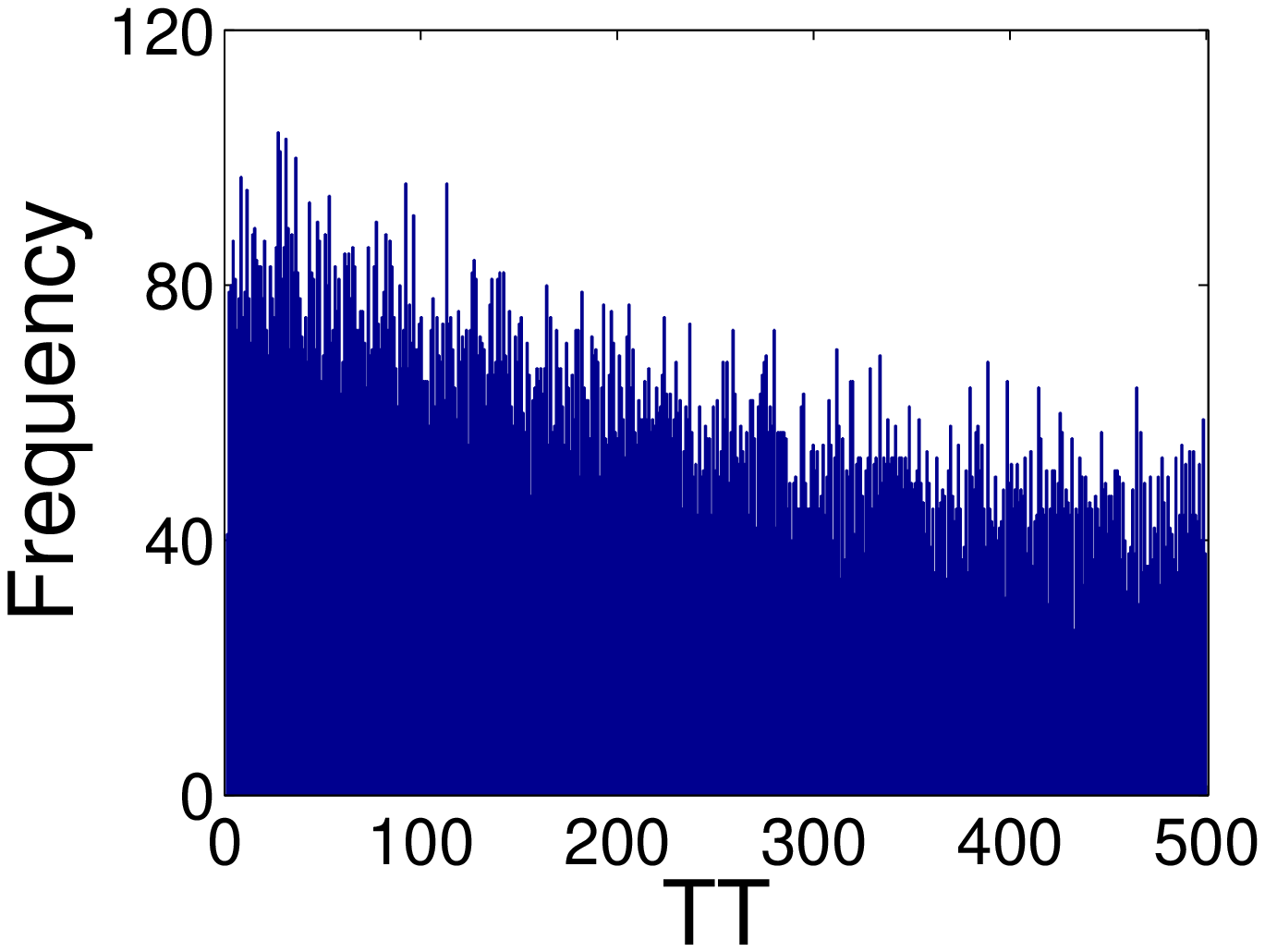}}
    \subfigure[$\lambda=9$]{\label{fig:csttlambda9}\includegraphics[scale=0.25]{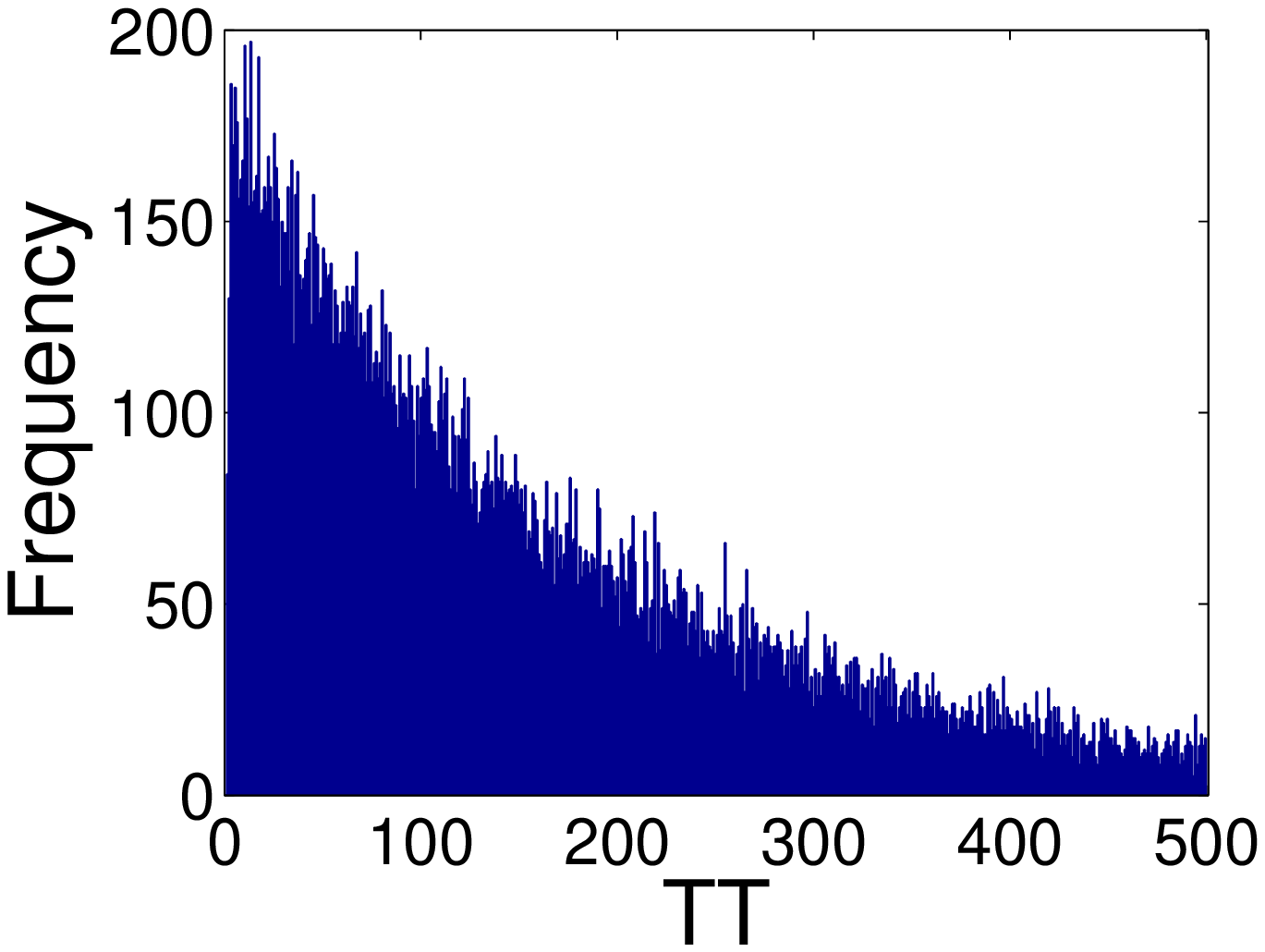}} \\
  \end{center}
  \caption{CS model distributions. TF distributions (top row); TT distributions (bottom row).}
  \label{fig:tttfcsdist}
\end{figure*}

The TF and TT distributions for the CS model on a lattice configuration are shown in Fig.~\ref{fig:tttfcsdist}. The CS model on a SF network demonstrates qualitatively similar distributions. Loadings $\lambda=6, 7$ correspond to transition loadings for these systems. The multi-modal nature of the TF distributions in Figs.~\ref{fig:cstflambda6},~\ref{fig:cstflambda7} indicate that multiple failure modes are present in the distributions. Multi-modal distributions have been observed in nature in the eruption of geysers \cite{azza:1} and sizes of ants \cite{weber:1}.

\begin{figure}[!]
  \begin{center}
    \subfigure[Multiple failures modes]{\label{fig:csmode}\includegraphics[scale=0.3]{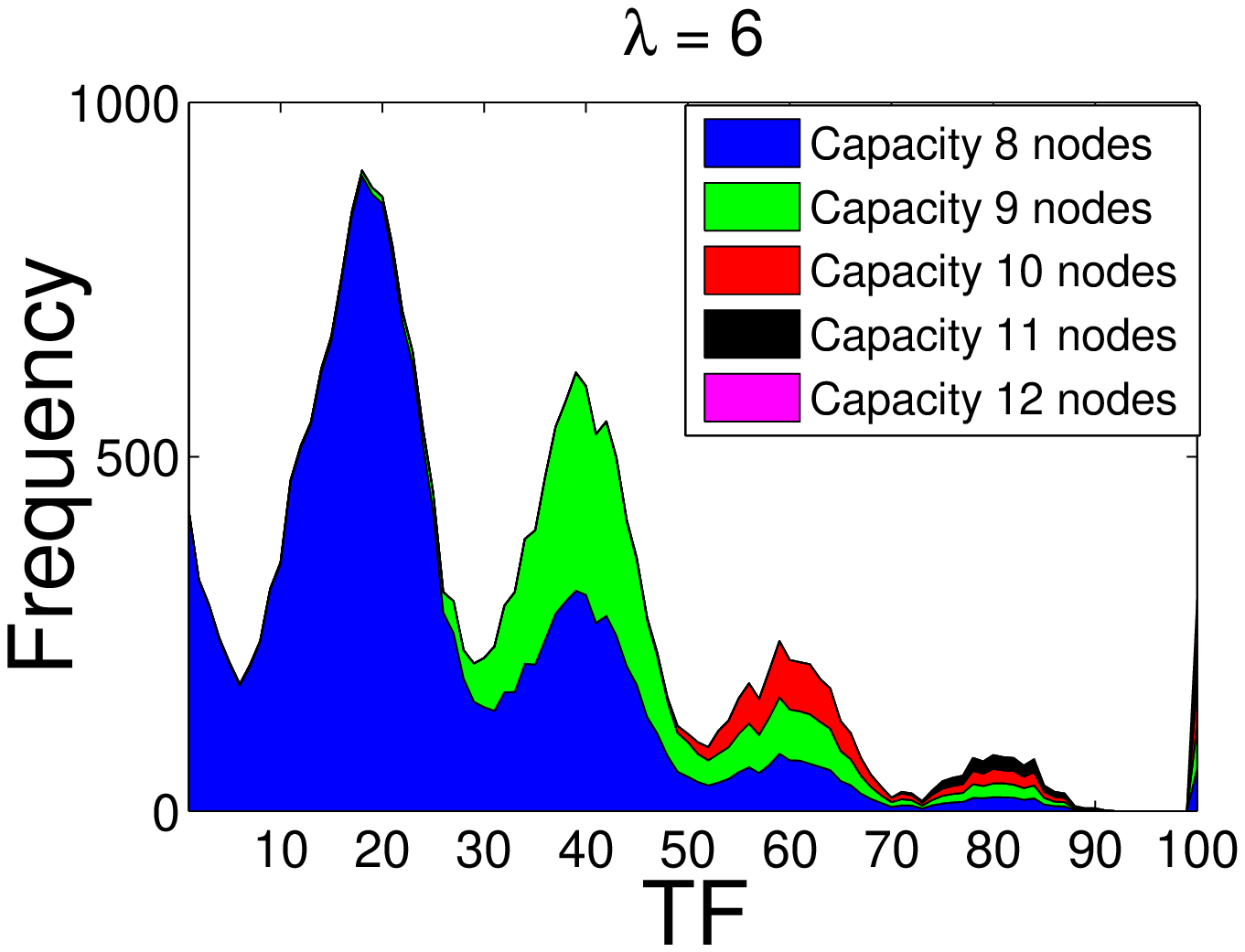}} \\
    \subfigure[Maximum Demand as a function of TF]{\label{fig:csdemand}\includegraphics[scale=0.3]{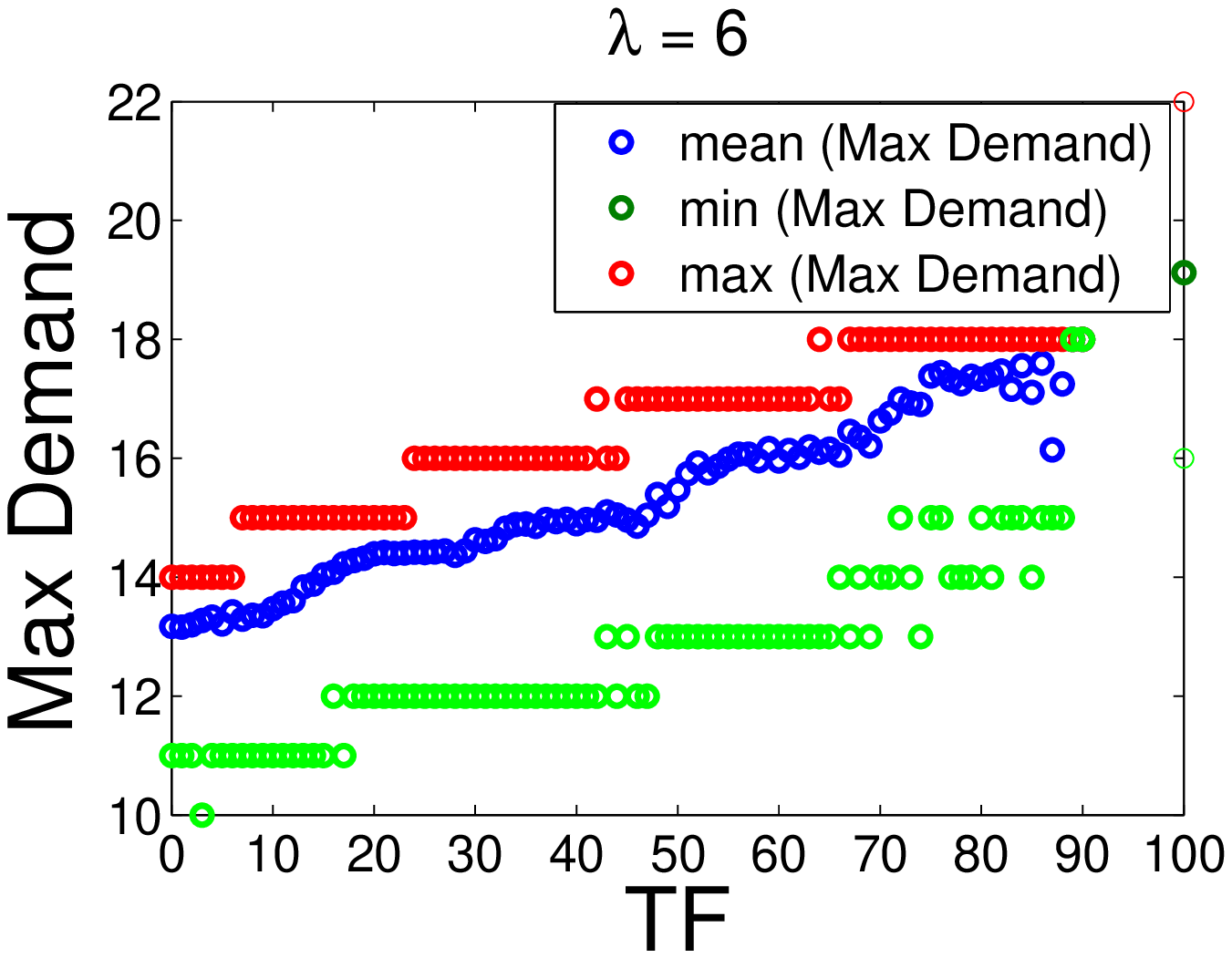}} \\
    \end{center}
  \caption{Extremal behavior of the CS model for $\lambda=6$ (Color Online)}
  \label{fig:failuremodescs}
\end{figure}

To isolate and identify the different failure modes, we filter the TF distributions based on the capacities of the failed components. Recall that in simulating the CS model we initialize component capacities from a integer uniform distribution $\mathcal{U}[8,12]$. In Fig.~\ref{fig:csmode} we color code the TF distribution for $\lambda=6$ [Fig.~\ref{fig:cstflambda6}] based on the capacities of the failed components. From Fig.~\ref{fig:csmode} the composition of the different failure modes becomes clear. The TF distribution for $\lambda=6$ is composed of a failure mode where only components of capacity$=8$ fail, a second failure mode where only components of capacity$=(8,9)$ fail, a third failure mode where only components of capacity$=(8,9,10)$ fail and so on. Similarly we can filter the TF distribution for $\lambda=7$.

It is also of interest to understand the dynamics that are exciting the multiple failure modes for transitions loadings $\lambda=6,7$. Motivated by Extreme Value Theory \cite{gumbel:1,*galambos:1} one explanation lies in the demand dynamics. Although the average demand on the system is $\lambda=6$ or $7$; the CS model is sensitive to the extremal behavior of the demand dynamics. Extremal events have been modeled in areas as diverse as finance \cite{embrechts:1} to earthquake characterization \cite{pisarenko:1}. In Fig.~\ref{fig:csdemand} we plot the extremal behavior of the demand dynamics as a function of TF for $\lambda=6$. The figure is constructed in the following way: in Fig.~\ref{fig:csmode} for each TF $\in [1, 100]$, we first determine the maximum demand seen by each of the systems in their associated window $[0, TT]$. For each TF $\in [1, 100]$ we then compute and plot, the mean maximum demand (shown in blue), the maximum maximum demand (shown in red) and the minimum maximum demand (shown in green).

In Fig.~\ref{fig:csdemand} we can clearly observe the staircase like growth trend of mean maximum demand as a function of TF and the step function growth of the associated min/max bounds of maximum demand. It is our opinion the extremal behavior of the demand dynamics \footnote{The number, size and sequence of extreme demand constitute the extremal behavior of the demand dynamics.} in conjunction with the structure of the component capacity topology is responsible for exciting the multiple distinct failure modes observed in Fig.~\ref{fig:csmode}.

For example in Fig.~\ref{fig:failuremodescs} consider the interval TF $\in [10, 25]$; mean maximum demand in this interval roughly corresponds to around 14 with components of capacity$=8$ failing. Noting that the queue size is 6, we can understand why components of capacity$=8$ are being overwhelmed by the mean maximum demand ($8+6=14$) in this interval. However, in addition to the specific sequence and number of extremal demands, relatively stronger neighborhood capacity topologies are partly responsible for the left side of the bell shape and relatively weaker neighborhood capacity topologies are partly responsible for the right side of the bell shape in the interval TF $\in [10, 25]$. For a specific level and sequence of extremal demand, a relatively stronger neighborhood capacity topology provides components a greater opportunity to survive through load sharing. We could construct similar arguments for the other bell curve like waves in Fig.~\ref{fig:failuremodescs} such as the interval TF $\in [30, 50]$ where components of capacity$=(8,9)$ are failing and mean maximum demand is approximately 15.

The result in Fig.~\ref{fig:failuremodescs} is similar in spirit to results in \cite{lvov:1}, where the authors show using shell models of turbulence that large fluid velocity fluctuations propagating from shell to shell cause multiscaling in the shell velocity variation distributions. In other words the velocity variation distribution is composed of two separate regions, the first due to ``small'' but ``usual'' velocity fluctuations and the second due to ``large'' but ``rare'' velocity fluctuations. Comparing to our results in Fig.~\ref{fig:failuremodescs}, we can see the extremal demand dynamics exciting different scales of failure in the TF distribution.

\begin{figure}[!]
  \begin{center}
    \subfigure[$\lambda=11$, TT distribution]{\label{fig:explambda11}\includegraphics[scale=0.25]{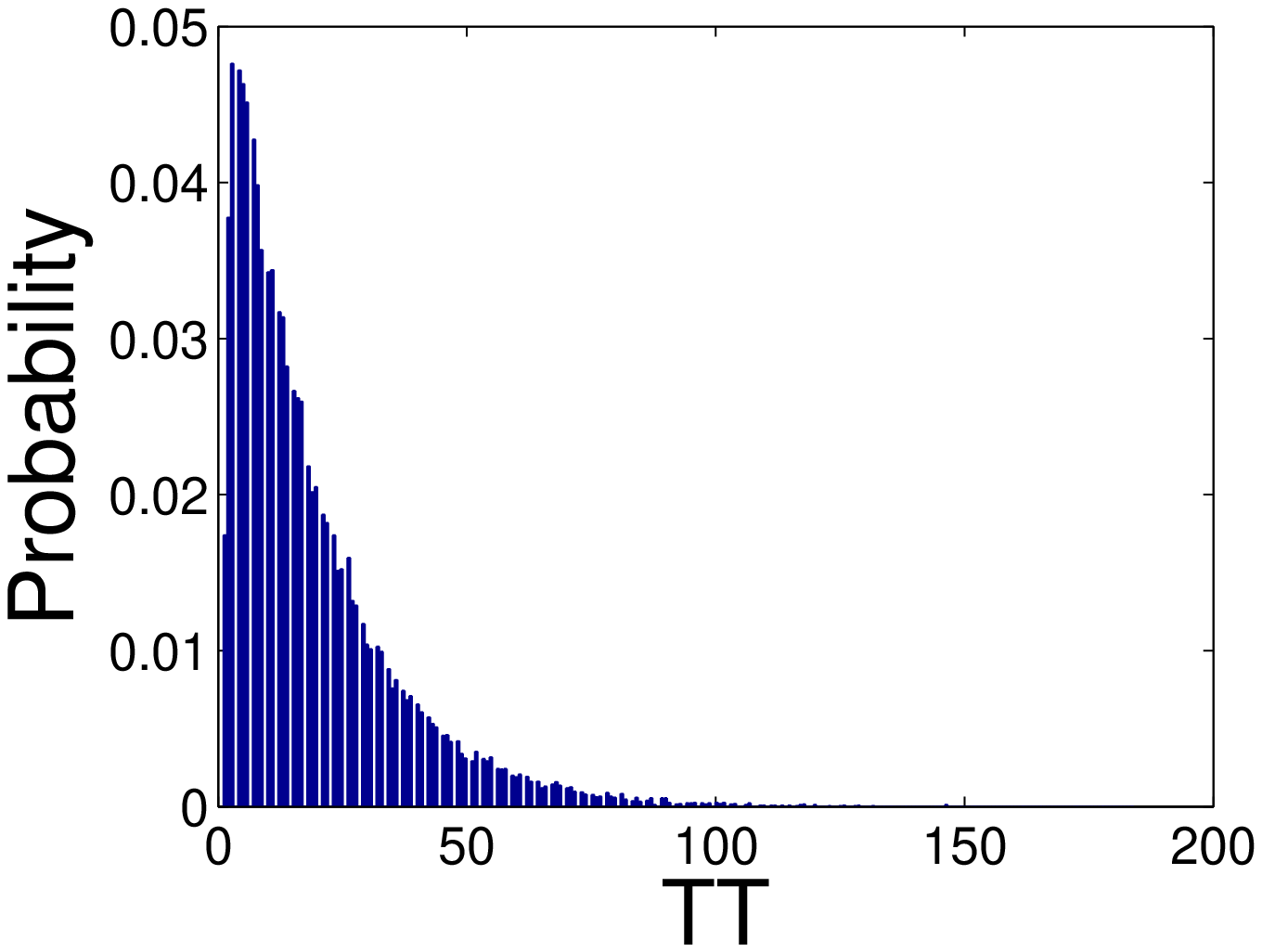}}
    \subfigure[$\lambda=11$, Fit to $P(TT)=Exp(\mu)$]{\label{fig:explambda11}\includegraphics[scale=0.25]{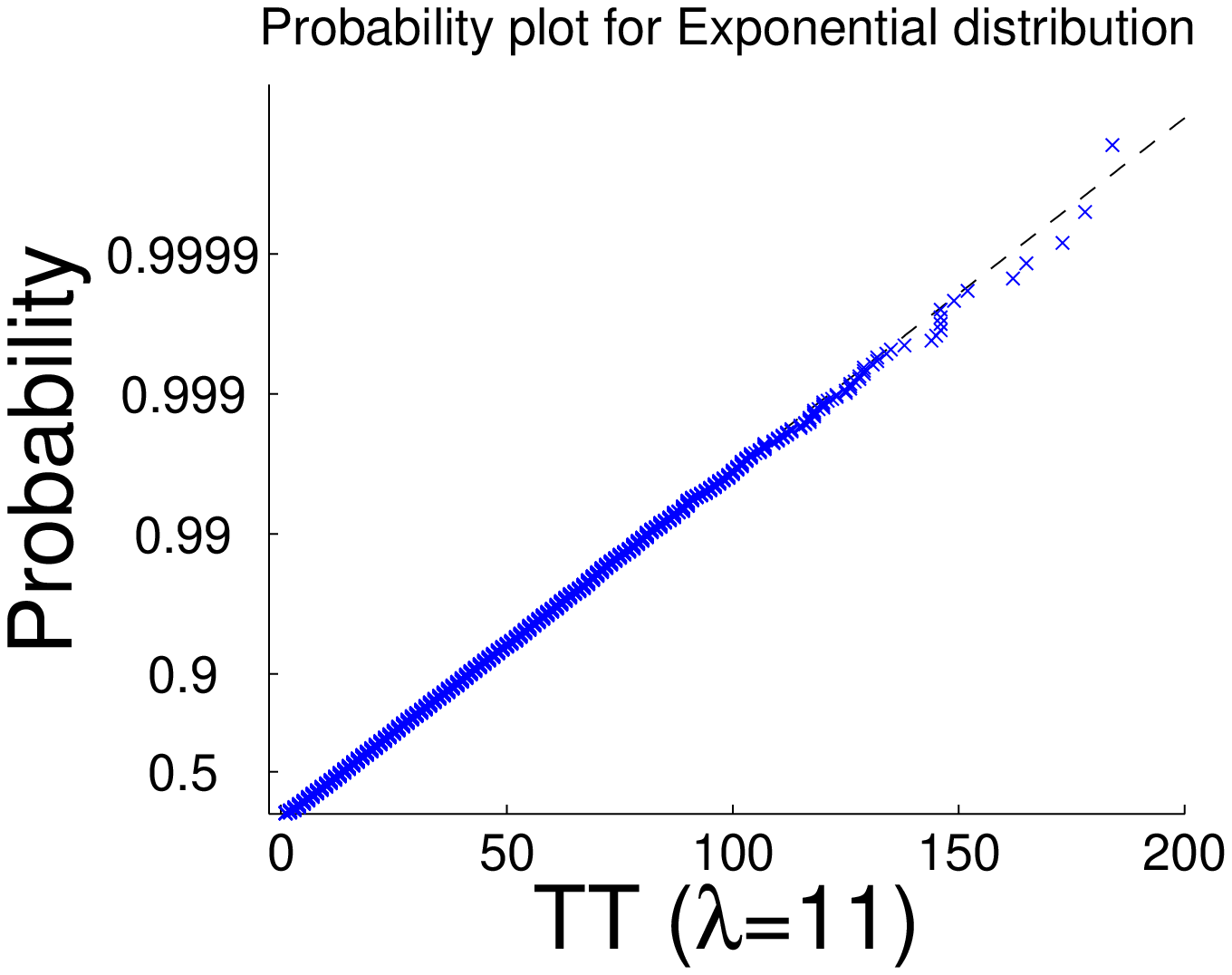}}
  \end{center}
  \caption{CS model TT distribution fit at the critical load}
  \label{fig:ttcomparecs}
\end{figure}

The TT distribution fit for the CS lattice model at the critical load $\lambda=11$ is shown in Fig.~\ref{fig:ttcomparecs}. At the critical load the CS model fits a exponential distribution $Prob(TT)=\frac{1}{\mu}e^{-TT/\mu}$ with parameter $\mu=19.06$ [Fig.~\ref{fig:explambda11}]. The CS model on SF networks demonstrates similar results. In communication and transportation applications M/M/1 queues have arrivals according to poisson processes and service time distributions are exponential \cite{bertsekas:1}. Although individual components in the CS model resemble M/D/1 queues \footnote{The interested reader will find an in-depth discussion on queueing theory and M/D/1, M/M/1 queues in particular in \cite{bertsekas:1}}, at critical demand rates the load sharing capability of the CS model is rendered redundant and the structure of the  component capacity topology causes the system to demonstrate exponential distribution failure times. Here we also note that exponential and sub-exponential distributions have been widely reported in financial applications such as drawdowns of the stock market, major currencies and major financial indices \cite{sornette:1,*johansen:2}. The relationship between the extremal dynamics of the CS model and market drawdowns presents an interesting subject for future investigation.

To summarize, we have used the concept of component strength and load interaction to investigate the failure mechanisms of complex systems utilizing two different strength/load interaction models. The LOS model explores strength/load interaction through a loss of strength mechanism. The CS model explores capacity/demand interaction through customer or data arrival/departure rate mechanism. At low levels of loading which correspond to lower network utilization, the failure mechanisms in the LOS model follow predictable trends [Eq.~\ref{eq:tttf} and Fig.~\ref{fig:tttfscale}]. The failures in the systems can be managed and network resources are sufficiently allocated. The system is resilient to cascading failure triggered by load redistribution.

At transition loadings or `tipping-points', both models demonstrate increasingly unpredictable behavior with system volatility and increasing disorder. Systems may or may not descend into catastrophic failure and extremal dynamics excite multiple failure modes in systems. The results imply that at these loadings the system resources (characterized by the system strength topology) need to be allocated appropriately to avoid catastrophic failure.

For critical loads system failure is reached through phase transitions. At criticality, depending on the strength/load interaction mechanism, systems demonstrate power law or exponential temporal failure patterns.

\begingroup
\squeezetable
\begin{table*}
\caption{\label{tab:ttfscale}Experimental values for the temporal system failure scaling phenomena of Equation.~\ref{eq:tttf}}
%\centering
\begin{tabular}{ | c | c | c | c | c | c | c | c | c | }
\hline
\multirow{5}{*} & $L$ & $\kappa_{1}$ & $\tau_{1}$ & Residual Norm & Breakpoint (TF) & $\kappa_{2}$ & $\tau_{2}$ & Residual Norm\\ \hline
{LOS Model} & 0.5 &  28.5 & 0.9 & 0.16 & - & - & - & -\\
{on a} & 1 & 46.4 & 0.66 & 0.16 & - & - & - & - \\
{Lattice network} & 1.5 & 61 & 0.7 & 0.13 & 25 & 428 & 0.05 & 0.28 \\ \hline
\multirow{5}{*} &  $L$ & $\kappa_{1}$ & $\tau_{1}$ & Residual Norm & Breakpoint (TF) & $\kappa_{2}$ & $\tau_{2}$ & Residual Norm\\ \hline
{LOS Model} & 0.5 &  21.54 & 0.83 & 0.29 & - & - & - & -\\
{on a} & 1 & 24.24 & 0.77 & 0.23 & - & - & - & - \\
{Scale-Free network} & 1.5 & 21.54 & 0.83 & 0.13 & 33 & 432.87 & 0.09 & 0.11 \\
{$\langle k \rangle = 12$} & 2 & 41.24 & 0.76 & 0.09 & 29 & 342.93 & 0.14 & 0.04 \\ \hline

\end{tabular}
\end{table*}
\endgroup

\end{document}